\documentclass[a4paper,12pt]{article}
\pdfoutput=1
\usepackage{graphicx,subfigure,amsmath,amssymb}
\usepackage{cite}
\usepackage{color}

\newlength{\dinwidth}
\newlength{\dinmargin}
 \def\la{ \langle}
  \def\ra{ \rangle}
 \def\r{ \gamma}
 \def\u{\mu}

 \def\lbd{\lambda}
 \def \d {{\rm d}}

\begin{document}

\title{
\begin{flushright}
{\large  SLAC-PUB-16849}\\[2mm] 
\end{flushright}
\bf Light-front holographic distribution amplitudes of pseudoscalar mesons and their application to  $B$-meson decays}
\author{Qin Chang$^{a,b}$, Stanley J. Brodsky$^{b}$ and Xin-Qiang Li$^{c}$\\
{ $^a$\small Institute of Particle and Nuclear Physics, Henan Normal University, Henan 453007,  China}\\
{ $^b$\small SLAC National Accelerator Laboratory, Stanford University, Stanford, CA 94309, USA}\\
{ $^c$\small Institute of Particle Physics and Key Laboratory of Quark and Lepton Physics~(MOE),} \\
{     \small Central China Normal University, Wuhan, Hubei 430079, China}}
\date{}
 \maketitle

\begin{abstract}\noindent
In this paper the dynamical spin effects of the light-front holographic wavefunctions for light pseudoscalar mesons are studied using  two different assumptions.
These improved wavefunctions are then confronted with a number of sensitive hadronic observables:  the decay constants of $\pi$ and $K$ mesons,  their $\xi$-moments, the pion-to-photon transition form factor, and the pure annihilation $\bar{B}_{s}\to \pi^+ \pi^-$ and $\bar{B}_{d}\to K^+K^-$ decays.
Taking $f_{\pi}$, $f_{K}$ and their ratio $f_{K}/f_{\pi}$ as constraint conditions, the $\chi^2$ analyses for holographic parameters, including the mass scale parameter $\sqrt{\lambda}$ and effective quark masses, are all consistent with the mass scale which controls the slopes of the light-quark hadronic Regge trajectories. In addition, we also show how the  improved light-front holographic distribution amplitudes regulate the end-point divergences which appear in the annihilation amplitudes of $B\to PP$ decays.
\end{abstract}


\newpage

\section{Introduction}

Light-front (LF) quantization is the natural frame-independent framework for the description of non-perturbative relativistic bound-state structure in quantum field theory. In principle, one can solve QCD by diagonalizing the LF QCD Hamiltonian $H_{LF}$, by using, for example, the discretized light-cone quantization method~\cite{Brodsky:1997de}. The spectrum and LF wavefunctions (LFWFs), which contain hadronic information, are then obtained from the eigenvalues and eigenfunctions of the Heisenberg equation $H_{LF}|\psi\rangle=M^2|\psi\rangle$. The result is an infinite set of coupled integral equations for the LF components in a Fock expansion~\cite{Brodsky:1997de}.  Unfortunately, solving these equations is a formidable computational task for the case of a non-abelian quantum field theory such as QCD in four-dimensional space-time. Consequently, alternative methods are necessary; for a recent comprehensive review, see Refs.~\cite{Brodsky:1997de,Brodsky:2014yha} for details.

In recent years, a semiclassical first approximation to strongly coupled QCD --  light-front holographic AdS/QCD  --
has been developed~\cite{deTeramond:2005su,Brodsky:2006uqa,Brodsky:2007hb, Brodsky:2008pf,deTeramond:2008ht}.  This color-confining approach predicts the spectroscopy of light-quark hadrons, dynamical observables such as form factors and structure functions, and the behavior of the running coupling in the nonperturbative domain.  Only one mass scale in addition to the quark masses appears.  This approach to hadron dynamics in physical four-dimensional spacetime at fixed LF time $\tau = x^+ = x^0 + x^3$   is holographically dual to the dynamics of a gravitational theory in five-dimensional anti-de Sitter~(AdS) space.
The LF eigenvalue equation can be reduced in this theoretical framework to an effective single-variable quantum-mechanical wave equation for $\phi(\zeta)$ which is given by~\cite{deTeramond:2008ht}
\begin{equation}\label{eq:scheq}
\left(-\frac{{\rm d}^2}{{\rm d}\zeta^2 }-\frac{1-4L^2}{4\zeta^2}+U(\zeta)\right) \phi(\zeta)=M^2\phi(\zeta)\,.
\end{equation}
The function  $U(\zeta)$ is the effective potential acting on the valence states~\cite{deTeramond:2010ge}; it is holographically related to a unique dilation profile in AdS space. As a result, one arrives at a concise form of a color-confining harmonic oscillator in impact space after the holographical mapping,  $U(\zeta,J)=\lambda^2\zeta^2+2\lambda(J-1)$.
The emergence of the mass scale $\lambda$ is consistent with the procedure of de Alfaro, Fubini, and Furlan~\cite{deAlfaro:1976vlx}
in which a mass scale can appear in a Hamiltonian without affecting the conformal invariance of the action\cite{Brodsky:2014yha}.

The eigenvalues of the light-front Schr\"odinger equation, Eq.~(\ref{eq:scheq}), are the squares of the meson masses.  The remarkably simple features of the empirical Regge trajectories for both meson and baryon families are reproduced by LF holographic QCD with only one parameter, the mass scale $\lambda$~\cite{deTeramond:2014asa,Dosch:2015nwa,Brodsky:2016yod,Brodsky:2016rvj,Dosch:2016zdv}. The predictions are in good agreement with the observed spectroscopy.
The eigensolutions of Eq.~(\ref{eq:scheq}) provide the $q \bar q$ light-front wavefunctions which control the dynamics of the mesons. After factoring out the longitudinal and orbital dependence, the LFWF can be written as
\begin{equation}
\psi(x,\zeta,\varphi)=e^{iL\varphi}X(x)\frac{\phi(\zeta)}{\sqrt{2\pi\zeta}}\,,
\end{equation}
where $\zeta^2=x(1-x){\bf b}_{\bot}^2$  is the Poincare' invariant radial variable of LF Hamiltonian, and  ${\bf b}_{\bot}$ is the invariant transverse impact variable.
The hadronic LFWF $\phi(\zeta)$  in the soft-wall holographic model encodes the dynamical properties of the mesons.
If one  also includes the light quark masses, it is given by~\cite{Brodsky:2007hb,Brodsky:2008pg}
\begin{equation} \label{eq:LFWFb}
\psi(x,\zeta)=\sqrt{ \frac{\lambda}{\pi}}\,\sqrt{x(1-x)}\,e^{-\frac{\lambda\zeta^2}{2}}e^{-\frac{1}{2\lambda}(\frac{m_q^2}{x}+\frac{m_{\bar{q}}^2}{1-x} )}
\end{equation}
in impact space.  Note that the LF  kinetic energy  $ \sum_i({k^2_\perp + m^2 \over x})_i$ is also the invariant mass squared ${\cal M}^2=(\sum_i k^\mu_i)^2$  of the hadronic constituents.

The holographic LFWF given by Eq.~(\ref{eq:LFWFb}) has been successfully used to describe diffractive $\rho$ meson electroproduction at HERA~\cite{Forshaw:2012im} as well as the spectroscopy and distribution amplitudes of light and heavy mesons~\cite{Branz:2010ub,Hwang:2012xf,Ahmady:2016ufq}. After introducing the LF spinor structure of the wavefunctions for light vector mesons in analogy with that of the photon, the authors of  Refs~\cite{Ahmady:2012dy,Ahmady:2013cva}  have predicted the light-front distribution amplitudes~(LFDAs) of the $\rho$ and $K^*$ vector mesons and have used them to evaluate the branching fractions of $B\to \rho\gamma$ and $B\to K^*\gamma$ decays.
In addition, the $B\to\rho\,,K^*$ form factors are computed and applied to rare $B \to K^* \mu^+ \mu^-$  and $B\to \rho \ell\bar{\nu}_{\ell}$
decays~\cite{Ahmady:2013cga,Ahmady:2014sva,Ahmady:2014cpa,Ahmady:2015gva,Ahmady:2015yea}.   The helicity dependence of the  LFWFs for the vector mesons is introduced in these analyses~\cite{Forshaw:2012im,Ahmady:2012dy,Ahmady:2013cva,Ahmady:2013cga,Ahmady:2014sva,Ahmady:2014cpa,Ahmady:2015gva,Ahmady:2015yea}  in order to predict specific helicity dependent observables. The helicity dependence of the holographic LFWF is assumed to decouple from the dynamics, which in turn leads to simple factorizable formulae for physical quantities, such as the decay constants.

In the past few years, several QCD-inspired approaches, such as QCD factorization (QCDF) \cite{Beneke1,Beneke2,Beneke3},  perturbative QCD (pQCD) \cite{KLS1,KLS2} and  soft-collinear effective theory (SCET) \cite{scet1,scet2,scet3,scet4}, have been developed  in order to evaluate the hadronic matrix elements of local operators which control two-body nonleptonic $B$ decays. However, the convolution integrals of the hard kernels with the asymptotic form of distribution amplitudes of light final states suffer from an end-point divergence, such as $ \int_0^1 du/u$ and $\int_0^1 du/(1-u)$. This divergence limits the prediction power of the theoretical approaches and prevents  reliable results.  At present, the dynamical origin of the end-point divergence is still unclear, even though there are conjectures about the reliability of the collinear approximation and concerns about our limited understanding of the QCD dynamics of hadrons.

In this paper, we will explore helicity-improved LFWFs for light pseudoscalar mesons, and then test their predictions for hadronic observables including  the decay constants of $\pi$ and $K$ mesons,  their $\xi$-moments and the pion-to-photon transition form factor.
We will also explore new applications to two-body nonleptonic $B$ decays, focusing especially on the measured pure annihilation $\bar{B}_s\to\pi^+\pi^-$ and $\bar{B}_d\to K^+K^-$ decay channels.  We will also show that the problem of end-point divergences in the annihilation amplitudes is mitigated by the improved behavior of the LFDAs near their end-points.

Several  schemes for regulating the end-point divergences have been previously proposed.  In the SCET approach,  a zero-bin subtraction~\cite{Manohar:2006nz} is assumed. The annihilation diagrams are found to be factorizable and bring no any strong phase in the leading terms of order ${\cal O}(\alpha_s(m_b)\Lambda_{\rm QCD}/m_b)$~\cite{Arnesen:2006vb}. In the QCDF approach, the end-point divergent integrals are treated as signals of infrared-sensitive contributions which can be regularized by introducing a complex quantity $X_A$~\cite{Beneke5,du2}.  Alternatively, one can introduce an infrared-finite dynamical gluon propagator which moves the end-point singularity into an integral over the time-like gluon momentum;  the divergence then vanishes, and a large strong phase is predicted~\cite{Chang:2008tf,Chang:2012xv}. In the pQCD approach, the end-point singularity is avoided by introducing parton transverse momentum $k_{T}$, but  at the expense of having to model the additional $k_{T}$ dependence of the meson distributions; this again predicts a large complex annihilation correction~\cite{KLS1,KLS2,Lu:2000em}.

In contrast, in the framework of the LF holographic QCD, the end-point contribution is naturally suppressed by the exponential factor in LFWF due to non-vanishing effective quark masses, $m_q$ and $m_{\bar{q}}$.
In this paper we will test if  the effective quark mass regulation of the end-point divergences in the annihilation amplitudes obtained from LF holographic QCD can provide viable predictions for pure annihilation heavy hadron decays.

Our paper is organized as follows. In section 2, the connections between holographic LFWFs and LFDAs for light pseudoscalar mesons are explored within the framework of LF quantization, based on two different assumptions for the helicity-dependence of the hadronic LFWFs.  Sections 3 and 4 are devoted to  numerical results and discussions in which the decay constants, the  $\xi$-moments and the pion-to-photon transition form factor are evaluated using the helicity-improved LFWFs and LFDAs. In section 5, the annihilation amplitudes and the pure annihilation $\bar{B}_{s}\to \pi^+ \pi^-$ and $\bar{B}_{d}\to K^+K^-$ decays are studied in detail  using the LFDAs predicted by   LF holographic QCD.  Finally, we give our summary in section 6.

\section{The holographic light-front wavefunctions and distribution amplitudes}

Our starting point is the definition of the distribution amplitudes (DAs) of light pseudoscalar meson~\cite{Brodsky:1997de,Lepage:1980fj}. The DAs parameterize the operator product expansion of meson-to-vacuum matrix elements~\cite{Braun:1989iv},
\begin{eqnarray}
\label{eq:twi2}
\la 0 | \bar q(0)\r_\u\r_5 q(x)|P(p)\ra&=&
  if_P p_\u \int_0^1 du \, e^{-i up\cdot x} \Phi(u)\,,\\
\label{eq:twi3a}
\la 0| \bar{q}(0)i\r_5 q(x) |P(p) \ra &=&
f_P\u_{P} \int_0^1 du\, e^{-i up\cdot x}  \phi(u)\,,
\end{eqnarray}
where $\mu_{P}=m_P^2/(\bar{m}_q+\bar{m}_{\bar{q}})$, $f_P$ is the decay constant of a pseudoscalar meson ($P$),  $\Phi(u)$ and $\phi(u)$ are twist-2 and twist-3 DAs, respectively.

In the following derivation, we will adopt Lepage-Brodsky~(LB) convention~\cite{Brodsky:1997de,Lepage:1980fj} and assume light-front gauge, $A^+=0$.
At equal LF time, the DAs can be expressed using Eqs.~(\ref{eq:twi2}) and (\ref{eq:twi3a}) as
\begin{eqnarray}
\label{eq:twi2p2}
f_ P\Phi(z, \u) &= &-\frac{i}{2}\int \d x^- e^{i z p^+x^-/2} \la 0 | \bar q(0)\r^+\r_5 q(x^-)| P(p)\ra\,,\\
\label{eq:twi3ap2}
\u_{ P}f_ P\phi(z, \u) &=&\frac{i}{ 2}\,p^+\int \d x^- e^{i z p^+x^-/2} \la 0| \bar{q}(0)\r_5 q(x^-) | P(p) \ra \,,
\end{eqnarray}
by performing the Fourier transformation with respect to $x^-=x^0-x^3$. The main remaining  task is to deal with the hadronic matrix elements in Eqs.~(\ref{eq:twi2p2}) and (\ref{eq:twi3ap2}).

In the framework of LF quantization~\cite{Lepage:1980fj,Brodsky:1997de}, a hadronic eigenstate $|P\ra$ can be expanded on a complete Fock-state basis of noninteracting 2-particle states as
\begin{equation}
|P\ra =\sqrt{4\pi N_c} \sum_{h,\bar{h}} \int \frac{\d k^+ \d^2{\bf k_{\bot}}}{(2\pi)^32\sqrt{k^+(p^+-k^+)}} \Psi^{P}_{h,\bar{h}}\left(k^+/p^+,{\bf k}_{\bot}\right)|k^+,k_{\bot},h;p^+-k^+,-k_{\bot},\bar{h}\ra \,,
\label{eq:Fockexp}
\end{equation}
in which, $\Psi^{P}_{h,\bar{h}}$ is the LFWF of the pseudoscalar meson with helicity-dependence included; $h$ and $\bar{h}$ are the helicities of quark and anti-quark, respectively; and the one-particle state is defined, for instance, by $|k^+\ra=\sqrt{2k^+}b^{\dag}|0\ra$. The Dirac (quark) field is expanded in terms of particle creation and annihilation operators as
\begin{equation}
\psi_+(x)= \int \frac{\d k^+}{\sqrt{2k^+}}\frac{ \d^2{\bf k}_{\bot}}{(2\pi)^3} \sum_h [b_{h} (k) u_h(k)e^{-ik\cdot x} + d^{\dagger}_{h} (k) v_{h} (k)e^{ik\cdot x}] \,,
\label{eq:qfexp}
\end{equation}
assuming  LF helicity spinors $ u_h$ and $v_{h}.$
The equal LF-time anti-commutation relations are
\begin{equation}
\{b^{\dagger}_{h} (k), b_{h'} (k^{\prime}) \}=\{d^{\dagger}_{h} (k), d_{h'} (k^{\prime}) \}= (2\pi)^3  \delta(k^+-k'^{+})\delta^2({\bf k}_{\bot}-{\bf k}'_{\bot}) \delta_{h h'}.
\label{anticommutation}
\end{equation}

Equipped with the above formulae, the hadronic matrix element in Eqs.~(\ref{eq:twi2p2}) and (\ref{eq:twi3ap2}) can be expressed as
\begin{eqnarray}
 \la 0 | \bar{q}(0) \Gamma q(x^-) |P(p) \ra&=&
\sqrt{4\pi N_c} \sum_{h,\bar{h}} \int \frac{\d k^{+}\d^2{\bf k}_{\bot}
\Theta(|\mathbf{k}_{\bot}| <\u)}
{(2\pi)^32\sqrt{k^+(p^{+}-k^{+})}}
\Psi^{P}_{h,\bar{h}}(k^+/p^+,{\bf k}_{\bot})\nonumber \\
&& \times
\bar{v}_{\bar{h}}(p^{+}-k^{+},-{\bf k}_{\bot}) \Gamma u_h(k^+,{\bf k}_{\bot})
e^{-ik^{+}x^{-}/2} \;,
\label{eq:meder1}
\end{eqnarray}
in which $\Gamma=\r^+\r_5$ and $\r_5$, and the scale $\mu$ is introduced as an ultraviolet cut-off on transverse momenta. Using Eq.~(\ref{eq:meder1}) and integrating over $x^-$ and $k^+$, we can further obtain a general expression for the RHS of Eqs.~(\ref{eq:twi2p2}) and (\ref{eq:twi3ap2}),
\begin{eqnarray}\label{eq:intleta}
\int \d x^- e^{izp^+x^-/2} \la 0 | \bar{q}(0)  \Gamma q(x^-)|P(p) \ra &=&
\frac{\sqrt{4\pi N_c}}{p^+}\sum_{h,\bar{h}} \int^{|\mathbf{k}_{\bot}| < \mu}
\frac{\d^2\mathbf{k}_{\bot}}{(2\pi)^3}\Psi^{P}_{h,\bar{h}}(z,\mathbf{k}_{\bot})\\ \nonumber
&&\times
\left \{ \frac{\bar{v}_{\bar{h}}((1-z)p^{+},-\mathbf{k}_{\bot})}{\sqrt{(1-z)}}
\Gamma \frac{u_h(zp^+,\mathbf{k}_{\bot})}{\sqrt{z}} \right \}\;.
\end{eqnarray}

To proceed with the derivation, we will need the explicit form of the holographic LFWF, $\Psi^{P}_{h,\bar{h}}$.  As mentioned in the introduction, the helicity-dependence of the holographic LFWF has been assumed in previous works to decouple from the dynamics,  and therefore $\Psi^{P}_{h,\bar{h}}(z,\mathbf{k}_{\bot})=\psi(z,\mathbf{k}_{\bot})$, where $\psi(z,\mathbf{k}_{\bot})$ is given by the Fourier transformation of Eq.~(\ref{eq:LFWFb}). This assumption leads to a universal formula for predicting physical quantities for different kinds of mesons; however, it is obviously disfavored by experiment.

In order to restore the proper helicity dependence, the holographic LFWF in the $\mathbf{k}_{\bot}$ space needs to be modified as
\begin{eqnarray}
\label{eq:LFWFP2}
\Psi_{h,\bar{h}}(z,\mathbf{k}_{\bot})&=&\frac{N}{\sqrt{4\pi}}
S_{h,\bar{h}}(z,\mathbf{k}_{\bot}) \psi(z,\mathbf{k}_{\bot}) \,,
\end{eqnarray}
where $S_{h,\bar{h}}(z,\mathbf{k}_{\bot})$ is the helicity-dependent wavefunction, $N$ is the normalization factor determined by the normalization condition
\begin{eqnarray}\label{eq:norc}
\sum_{h,\bar{h}}\int \d z \frac{\d^2\mathbf{k}_{\bot}}{(2\pi)^2} |\Psi_{h,\bar{h}}(z,\mathbf{k}_{\bot})|^2
=\sum_{h,\bar{h}}\int \d z \d^2\mathbf{b}_{\bot} |\Psi_{h,\bar{h}}(z,\mathbf{b}_{\bot})|^2
=1\,,
\end{eqnarray}
and $\psi(z,\mathbf{k}_{\bot})$ is  the radial wavefunction obtained by performing the Fourier transformation of Eq.~(\ref{eq:LFWFb}),
 \begin{eqnarray}
\label{eq:LFWFkT}
\psi(z,\mathbf{k}_{\bot})=\,\frac{4\pi}{\sqrt{\lbd}}\frac{1}{ \sqrt{z(1-z)}}\,e^{-\frac{\mathbf{k}_{\bot}^2}{2\lbd\,z(1-z)}}e^{-\frac{1}{2\lbd}(\frac{m_q^2}{z}+\frac{m_{\bar{q}}^2}{1-z} )}\,.
 \end{eqnarray}

In the case of a vector meson, one can work in analogy with the lowest-order helicity structure of the photon LFWF in QED;  the following structure of $S^V_{h,\bar{h}}$ is thus assumed~\cite{Forshaw:2012im}
\begin{eqnarray}
\label{eq:LFWFsV2}
S_{h,\bar{h}}^{V,\lbd}(z,\mathbf{k}_{\bot})=
\bar{u}_{h} (zp^+ ,\mathbf{k}_{\bot} )\not\!\epsilon^{\lbd}v_{\bar{h}}((1-z)p^+,-\mathbf{k}_{\bot})\,.
\end{eqnarray}
This form has been successfully used to study the production of the $\rho$ and $K^*$ vector mesons and the decays related to $B\to\rho, K^*$  transitions~\cite{Ahmady:2013cga,Ahmady:2014sva}.

In the case of a pseudoscalar meson, following such a strategy, $\not\!\epsilon^{\lbd}$ in Eq.~(\ref{eq:LFWFsV2}) would be replaced simply by $\r_5$~\cite{Geng:2016pyr,Jaus:1989au,Choi:2007yu}. Very recently, this spin structure has been used to evaluate the holographic DA of $\pi$ meson  in Ref.~\cite{Ahmady:2016ufq}.
The helicity-dependent wavefunction is written explicitly as
\begin{eqnarray}
\label{eq:LFWFsP2}
S_{h,\bar{h}}^{P}(z,\mathbf{k}_{\bot})=
\bar{u}_{h} (zp^+ ,\mathbf{k}_{\bot} )(i\r_5)v_{\bar{h}}(\bar{z}p^+,-\mathbf{k}_{\bot})\,, \qquad {\rm Scenario~1}\,
\end{eqnarray}
where the factor ``$i$'' is now added to be consistent with the convention for the definition in Eqs.~(\ref{eq:twi2}) and (\ref{eq:twi3a}), and the abbreviation $\bar{z}=1-z$ is used for convenience.  An additional multiplying factor ``$M_P$'', which is added in Ref.~\cite{Ahmady:2016ufq}, has been absorbed into the normalization constant.
It should be noted, however, that this spin structure corresponds to light quark and antiquark of the pseudoscalar meson, such as the pion, to have  parallel spin projections, and thus $L^z = \pm 1$. This state has twist$=2+L=3$, and it is thus not the meson eigenstate of the AdS/QCD theory.
Instead of $\r_5$, the Dirac structure like $\not\!p\r_5$ is also allowed.  We therefore consider an alternative form of $S_{h,\bar{h}}^{P}$:
\begin{eqnarray}
\label{eq:LFWFsP3}
S_{h,\bar{h}}^{P}(z,\mathbf{k}_{\bot})=
\bar{u}_{h} (zp^+ ,\mathbf{k}_{\bot} )(i\frac{\widetilde{m}_P}{2p^+}\r^+\r_5+i\r_5)v_{\bar{h}}(\bar{z}p^+,-\mathbf{k}_{\bot})\,, \qquad {\rm Scenario~2}\,
\end{eqnarray}
in which, the structure $\r^+\r_5$ implies that the light quark and antiquark have only opposite helicities.    This is the helicity assignment that couples the pion to the axial-vector current and thus pion decay constant $f_\pi$ in: $ \pi^- \to W^- \to \ell^- {\bar\nu}.$
It is thus the leading-twist LFWF,  and is the solution from AdS/QCD for light quarks.
Since $\widetilde{m}_P$ is the invariant mass of $q\bar{q}$ pair in the $P$ meson,  the  dimensions of the two terms in $S_{h,\bar{h}}^{P}$, Eq.~(\ref{eq:LFWFsP3}), are consistent.

In the following, for convenience of discussion, the two helicity-dependent wavefunctions defined by Eqs.~\eqref{eq:LFWFsP2} and  (\ref{eq:LFWFsP3}) will be referred to as Scenario~1~(S1) and Scenario~2~(S2), respectively.  They are related by the Gell-Mann-Oakes-Renner (GMOR) relation and are thus not independent~\cite{Brodsky:2012ku}.
Using LB convention~\cite{Lepage:1980fj}, the two helicity-dependent wavefunctions $S_{h,\bar{h}}^{P}$ are given explicitly as
\begin{eqnarray}
S_{h,\bar{h}}^{P}(z,\mathbf{k}_{\bot})=
 \left\{\begin{aligned}
\frac{i}{\sqrt{z\bar{z}}}\left[ -|\mathbf{k}_{\bot}|e^{\mp i \theta_k}\delta_{h\pm,\bar{h}\pm}\pm\left(zm_{\bar{q}}+\bar{z}m_{q}\right) \delta_{h\pm,\bar{h}\mp}\right]\,,\qquad \qquad~{\rm Scenario~1} \\
\frac{i}{\sqrt{z\bar{z}}}\left[  -|\mathbf{k}_{\bot}|e^{\mp i \theta_k}\delta_{h\pm,\bar{h}\pm}\pm\left(zm_{\bar{q}}+\bar{z}m_{q}+z\bar{z}\widetilde{m}_P\right) \delta_{h\pm,\bar{h}\mp}\right]\,,\quad {\rm Scenario~2}
\end{aligned}\right.
\end{eqnarray}
and the spinor currents in Eq.~\eqref{eq:intleta} can be written as
\begin{eqnarray}
\label{eq:DGk1}
&&\frac{\bar{v}_{\bar{h}}}{\sqrt{\bar{z}}} \r^+\r_5 \frac{u_h}{\sqrt{z}}=\pm\, 2p^+ \delta_{h\pm,\bar{h}\mp}\,,\\
\label{eq:DGk2}
&&\frac{\bar{v}_{\bar{h}}}{\sqrt{\bar{z}}} \r_5 \frac{u_h}{\sqrt{z}}=\frac{1}{z\bar{z}}\left[|\mathbf{k}_{\bot}|e^{\pm i \theta_k}\delta_{h\pm,\bar{h}\pm}  \mp\left(zm_{\bar{q}}+\bar{z}m_{q}\right) \delta_{h\pm,\bar{h}\mp} \right]\,,
\end{eqnarray}
in which, $\mathbf{k}_{\bot}=|\mathbf{k}_{\bot}|e^{\pm i \theta_k}$ is specified.

Finally, in the $\mathbf{k}_{\bot}$ space, using the building blocks given above, the holographic DAs of $P$ meson
can be written as
\begin{eqnarray}
 \label{eq:lfdak2}
\Phi(z, \u) [{\rm S1}] &=& \frac{\sqrt{N_c }}{\pi f_P}  \int^{|\mathbf{k}| < \mu}
\frac{\d^2\mathbf{k}_{\bot}}{(2\pi)^2}\,  \frac{N_1}{(z\bar{z})^{1/2}}(\bar{z}m_q+zm_{\bar q})\,\psi(z,\mathbf{k}_{\bot})\,,\\
 \label{eq:lfdak3p}
\phi(z, \u)[{\rm S1}] &=& \frac{ \sqrt{N_c } }{ 2\pi \u_{P}f_P}  \int^{|\mathbf{k}| < \mu}
\frac{\d^2\mathbf{k}_{\bot}}{(2\pi)^2}\frac{N_1}{(z\bar{z})^{3/2}}\Big\{\mathbf{k}_{\bot}^2+ (zm_{\bar{q}}+\bar{z}m_q)^2  \Big\} \,\psi(z,\mathbf{k}_{\bot})\,,
\end{eqnarray}
in the case of S1~\footnote{Very recently, in Ref.~\cite{Ahmady:2016ufq}, the twist-2 holographic LFDA of $\pi$ meson is also evaluated with a $S_{h,\bar{h}}^{P}$ similar to S1.
Our result is  more general  in comparison  with that of Ref.~\cite{Ahmady:2016ufq}.}, and
\begin{align}
 \label{eq:lfdak22}
\Phi(z, \u) [{\rm S2}] &= \frac{\sqrt{N_c } }{\pi f_P  }  \int^{|\mathbf{k}| < \mu}
\frac{\d^2\mathbf{k}_{\bot}}{(2\pi)^2}\,  \frac{N_2}{(z\bar{z})^{1/2}}(\bar{z}m_q+zm_{\bar q}+z\bar{z}\widetilde{m}_P)\,\psi(z,\mathbf{k}_{\bot})\,,\\
 \phi(z, \u)[{\rm S2}] &= \frac{ \sqrt{N_c }}{ 2\pi \u_{P}f_P}  \int^{|\mathbf{k}| < \mu}
\frac{\d^2\mathbf{k}_{\bot}}{(2\pi)^2}\frac{N_2}{(z\bar{z})^{3/2}}\Big\{\mathbf{k}_{\bot}^2
+ (zm_{\bar{q}}+\bar{z}m_q)(zm_{\bar{q}}+\bar{z}m_q+z\bar{z}\widetilde{m}_P )  \Big\} \,\psi(z,\mathbf{k}_{\bot})\,, \label{eq:lfdak3p2}
\end{align}
in the case of S2, where $N_1$ and $N_2$ are the corresponding normalization factors determined by Eq.~(\ref{eq:norc}).
The expression in the impact space can be obtained through Fourier transformation. These formulae, which exhibit the connections between holographic LFDAs and LFWFs, are one of the main theoretical results in this paper. Using the theoretical framework given above, we will present numerical results and applications of these holographic LFDAs and LFWFs in the following sections.


\section{Input parameters and decay constants}

\subsection{Inputs}

Before presenting our numerical results, we now clarify the values of input parameters used in our evaluation. One of the most important inputs is the mass scale parameter $\sqrt{\lbd}$~\footnote{In some references, the parameter $\kappa=\sqrt{\lbd}$ is used.}, which could be extracted from many observables. For example, to fit the light-quark mass spectrum, the values $\sqrt{\lbd}=0.59\,{\rm GeV}$ and $0.54\,{\rm GeV}$ are suggested in Ref.~\cite{Brodsky:2014yha} for light pseudoscalar and vector mesons, respectively. A mean value, $\sqrt{\lbd}=0.523\,{\rm GeV}$, is obtained in Ref.~\cite{Brodsky:2016rvj} by fitting all of the slopes of the different Regge trajectories for mesons and baryons including all excitations. This result is also favored by the recent high accuracy computation of the perturbative QCD scale parameter $\Lambda_{\overline{\rm MS}}$~\cite{Deur:2016opc}. The fit to the Bjorken sum-rule data at low $Q^2$ yields $\sqrt{\lbd}= 0.496 \pm 0.007\,{\rm GeV}$~\cite{Prosperi:2006hx}. In Ref.~\cite{Deur:2016cxb}, the value  $\sqrt{\lbd}= 0.51\pm 0.04\,{\rm GeV}$ is used for determining the freezing value of $\alpha_s(Q^2)$ and the interface between perturbative and nonperturbative QCD.  In addition, in order to describe the HERA data on diffractive $\rho$ and $\phi$ electroproduction, the values $\sqrt{\lbd}= 0.55\,{\rm GeV}$ and $0.56\,{\rm GeV}$ are suggested~\cite{Forshaw:2012im,Ahmady:2016ujw}. Besides $\sqrt{\lbd}$, the  light-quark masses appearing in the holographic LFWFs are the other important inputs, which will be specified below.

In this paper, for S1, we follow entirely the inputs suggested by the recent study of holographic DA of $\pi$ meson with a similar LFWF of S1~\cite{Ahmady:2016ufq}. Explicitly, the following input values are used~\cite{Ahmady:2016ufq}:
\begin{equation}\label{eq:input1}
\sqrt{\lbd}=523~{\rm MeV}\,,\quad m_s=450 ~{\rm MeV}\,,  \quad m_{u,d}=330~{\rm MeV}\,,\quad {\rm Scenario~1}
\end{equation}
where the constituent quark masses are adopted, and are also used for studying the $\rho$ and $K^*$ mesons~\cite{Ahmady:2013cga,Ahmady:2014sva}. It should be noted that, as pointed out in Ref.~\cite{Brodsky:2014yha}, the light-quark masses introduced in the holographic LFWF are not the traditional constituent masses in the non-relativistic theories, but are the effective quark masses from the reduction of higher Fock states as functionals of the valence states. Such effective quark masses, in principle, should be universal in a specific theoretical framework of holographic QCD.

For S2, on the other hand, we take
\begin{equation}\label{eq:input2}
\sqrt{\lbd}=590\pm15\,{\rm MeV}\,,\quad m_s=272^{+69}_{-37} \,{\rm MeV}\,,  \quad m_{u,d}=79^{+7}_{-5}\,{\rm MeV}\,,\quad {\rm Scenario~2}
\end{equation}
which will be further explained in detail in the next subsection. It should be noted that such input values are very similar to the results~\cite{Brodsky:2014yha},
\begin{equation}\label{eq:regtraj}
\sqrt{\lbd}=590\,{\rm MeV}\,, \quad m_s=357 \,{\rm MeV}\,, \quad  m_{u,d}=46\,{\rm MeV}\,,
\end{equation}
obtained by fitting the Regge trajectories of pseudoscalar mesons in the framework of LF holographic QCD~\cite{Brodsky:2014yha}.

\subsection{Decay constants}

\begin{table}[t]
\caption{\label{tab:dc} Numerical results of decay constants of $\pi^-$ and $K^-$ mesons in unit of ${\rm MeV}$.}
\let\oldarraystretch=\arraystretch
\renewcommand*{\arraystretch}{1.1}
\begin{center}\setlength{\tabcolsep}{1.0pt}
\begin{tabular}{lccccccccc}
\hline\hline
&Exp.     &S1&S2& ETM & HPQCD& FL/MILC & LQCD Ave.\\
& \cite{PDG}&  &  &  \cite{Carrasco:2014poa}  & \cite{Dowdall:2013rya} & \cite{Bazavov:2014wgs} & \cite{PDG,Aoki:2016frl}\\ \hline
$f_{\pi}$& $130.28\pm0.26$ &$132.84$ &$130.10^{+3.23}_{-3.77}$&  --- &$130.39\pm0.20$&  ---&$130.2\pm1.7$\\
$f_{K}$ & $156.09\pm0.49$ &  $136.04$  &$156.04^{+5.09}_{-4.45}$ & $154.1\pm2.1$&$155.37\pm0.34$ & $155.92^{+0.43}_{-0.36}$&$155.6\pm0.4$ \\
$\frac{f_{K}}{f_{\pi}}$ & $1.198\pm0.004$ &$1.024$ &$1.199^{+0.032}_{-0.030}$ &$1.184\pm0.016$ &$1.1916\pm0.0022$& $1.1956^{+0.0028}_{-0.0021}$&$1.1928\pm0.0026$\\
\hline\hline
\end{tabular}
\end{center}
\end{table}

The values of holographic parameters are constrained by the decay constants. So, first, we present our predictions for the decay constant of pseudoscalar meson, which is defined as
\begin{eqnarray}
 \la 0 | \bar q\r^\u\r_5 q|P(p)\ra=if_P p^{\u}\,.
\end{eqnarray}
Expanding the hadronic state in the same manner as in section 2, we can finally arrive at
\begin{eqnarray}
\label{eq:dck1}
f_P&=&\frac{\sqrt{N_c}}{\pi}\int_0^1\d z\int\frac{\d^2{\bf k}_{\bot}}{(2\pi)^2} \frac{\bar{z}m_q+zm_{\bar{q}}}{\sqrt{z\bar{z}}}N_1\psi(z,\mathbf{k}_{\bot}) \,, \quad {\rm Scenario~1}\\[0.2cm]
\label{eq:dck2}
f_P&=&\frac{\sqrt{N_c}}{\pi}\int_0^1\d z\int\frac{\d^2{\bf k}_{\bot}}{(2\pi)^2} \frac{\bar{z}m_q+zm_{\bar{q}}+\bar{z}z\widetilde{m}_P}{\sqrt{z\bar{z}}}N_2\psi(z,\mathbf{k}_{\bot}) \,. \quad {\rm Scenario~2}
\end{eqnarray}

With the inputs mentioned above, our numerical results for $f_{\pi}$, $f_{K}$ and their ratio $ f_{K}/f_{\pi}$ are summarized in Table~\ref{tab:dc}, in which the theoretical errors in S2 are obtained by evaluating separately the uncertainties induced by each input parameter in Eq.~\eqref{eq:input2} and then adding them in quadrature. For comparison, the latest experimental data~\cite{PDG}~\footnote{ The values $|V_{ud}|=0.9758\pm 0.0016$ and $|V_{us}|=0.2248\pm0.0006$~\cite{PDG} are used to obtain the data of $f_{\pi}$ and $f_{K}$.},  the recent results based on lattice QCD~(LQCD) with $N_f=2+1+1$ obtained by ETM~\cite{Carrasco:2014poa}, HPQCD~\cite{Dowdall:2013rya}, Fermilab Lattice~(FL) and MILC Collaborations~\cite{Bazavov:2014wgs}, and the world averaged results of LQCD~\cite{PDG,Aoki:2016frl} are also listed in Table~\ref{tab:dc}.

In S1, our result $f_{\pi}=132.84\,{\rm MeV}$ is comparable with the data and, as found in Ref.~\cite{Ahmady:2016ufq}, achieves a much better agreement than the result without helicity improvement. However, S1 results in very small results for $f_K=136.04\,{\rm MeV}$ and $f_K/f_{\pi}=1.024$, which deviates far from the data. In fact, no matter what values of the light-quark masses are used, the predicted $f_K/f_{\pi}$ in S1 is always much smaller than the data and the LQCD results. This implies that S1 cannot provide sufficient flavor-asymmetry resources. It is very interesting to note that this deficiency in S1 can be remarkably improved in S2. From Table~\ref{tab:dc}, it can be seen that all the results in S2 are in good agreement with the data and the LQCD results.

\begin{figure}[t]
\begin{center}
\subfigure[]{\includegraphics[scale=0.64]{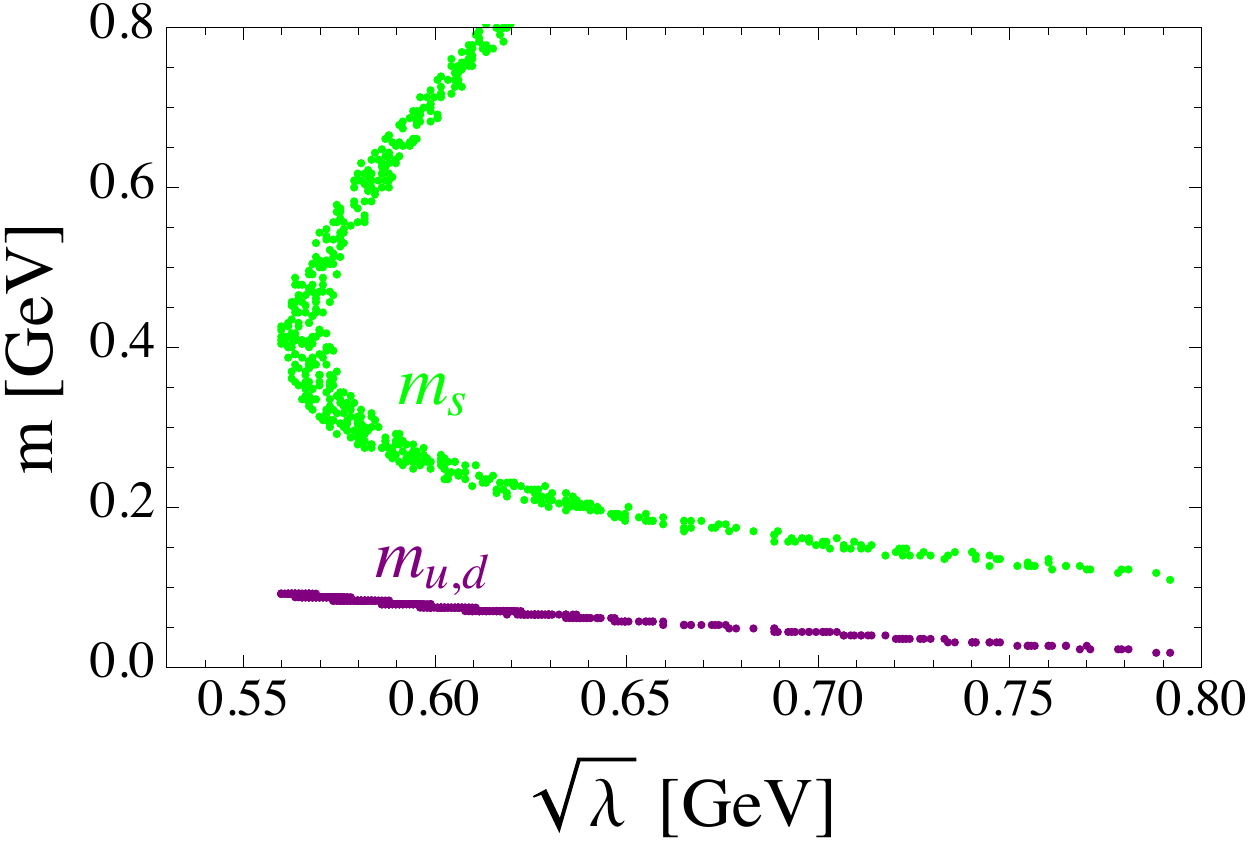}}\quad
\subfigure[]{\includegraphics[scale=0.65]{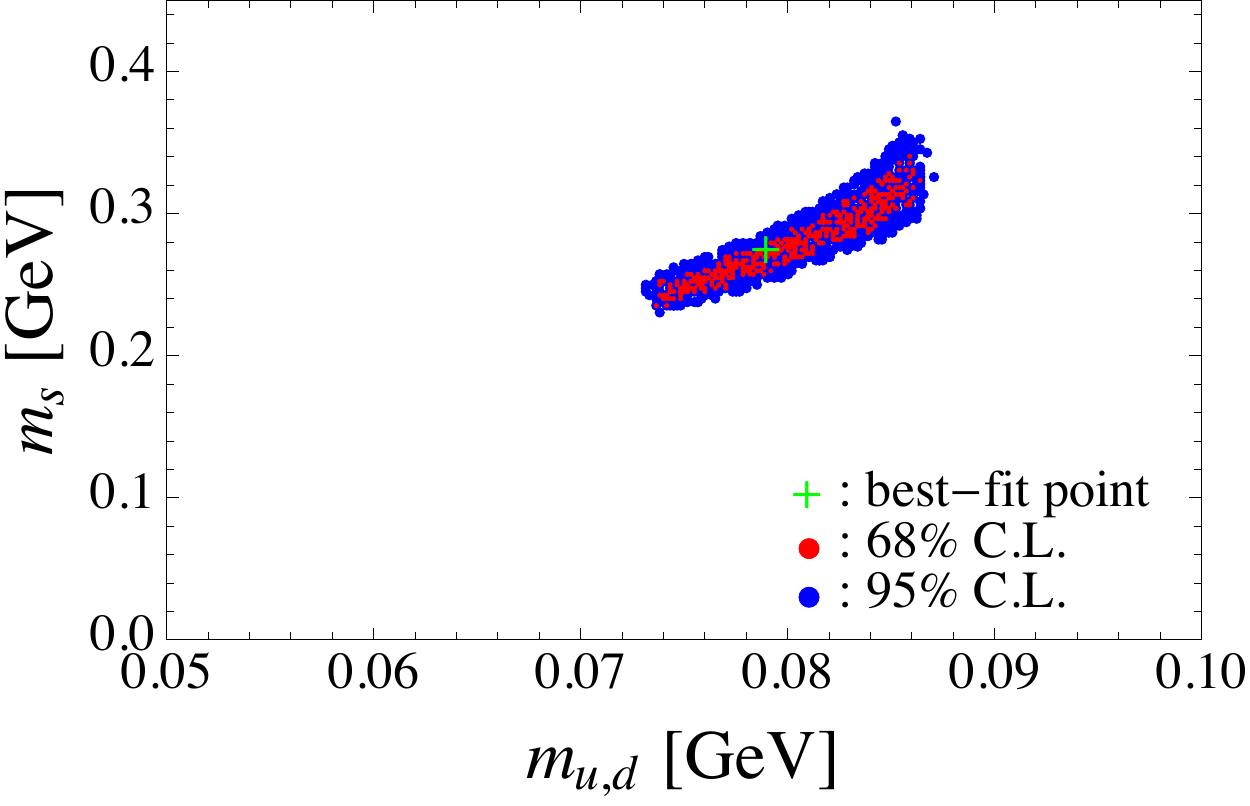}}
\caption{\label{fig:fitspac}The fitted spaces for the holographic parameters in S2 under the constraints from the decay constants $f_{\pi}$ and $f_{K}$ and their ratio $f_K/f_{\pi}$. Fig.~(a): the allowed spaces of $\sqrt{\lbd}$, $m_s$ and $m_{u,d}$ at $95\%$ C.L.; Fig.~(b): the allowed spaces of $m_s$ and $m_{u,d}$ with $\sqrt{\lbd}=0.590\pm0.015 ~{\rm GeV}$.}
\end{center}
\end{figure}

The decay constants $f_{\pi}$ and $f_{K}$ are very sensitive to the holographic parameters, $\sqrt{\lbd}$, $m_s$ and $m_{u,d}$, and we can, therefore, perform a $\chi^2$-fit for these parameters using the experimental data on $f_{\pi}$, $f_K$ and $f_K/f_{\pi}$ listed in Table~\ref{tab:dc}.  Our fitting results for $\sqrt{\lbd}$, $m_s$ and $m_{u,d}$ at $95\%$ C.L. are shown in Fig.~\ref{fig:fitspac}~(a). Even though the parameter spaces could not be seriously constrained due to the limited constraining conditions, we do obtain some useful bounds, $m_s\gtrsim100~{\rm MeV}$, $m_{u,d}\lesssim100~{\rm MeV}$ and $\sqrt{\lbd}>550~{\rm MeV}$. The bound $\sqrt{\lbd}>550~{\rm MeV}$ confirms the finding in Ref.~\cite{Brodsky:2014yha} that a relatively larger $\sqrt{\lbd}\sim 590~{\rm MeV}$ for pseudoscalar mesons is required compared with $\sqrt{\lbd}\sim 540~{\rm MeV}$ for vector mesons. Thus, in our evaluation, we take the value $\sqrt{\lbd}=590\,{\rm MeV}$ and assign a conservative uncertainty $\pm15\,{\rm MeV}$.

With $\sqrt{\lbd}$ fixed at $\sqrt{\lbd}=590\pm15\,{\rm MeV}$, our fitted results for $m_s$ and $m_{u,d}$ are shown in Fig.~\ref{fig:fitspac}~(b), and the corresponding numerical results are given by Eq.~\eqref{eq:input2};  another solution with unacceptably large $m_s\sim 700\, {\rm MeV}$, which is allowed in principle (see Fig.~\ref{fig:fitspac}~(a)), is discarded. It can be seen from Fig.~\ref{fig:fitspac}~(b) that the allowed spaces are strongly constrained. Comparing Eqs.~\eqref{eq:input2} with \eqref{eq:regtraj}, we note that the fitted results for the holographic parameters match the parameters obtained by fitting the Regge trajectories of pseudoscalar mesons~\cite{Brodsky:2014yha} (the small difference is acceptable due to the modified LFWFs used in this paper).

\section{Holographic DAs and pion-to-photon form-factor}

\subsection{The results of holographic DAs}

\begin{figure}[t]
\begin{center}
\subfigure[]{\includegraphics[scale=0.65]{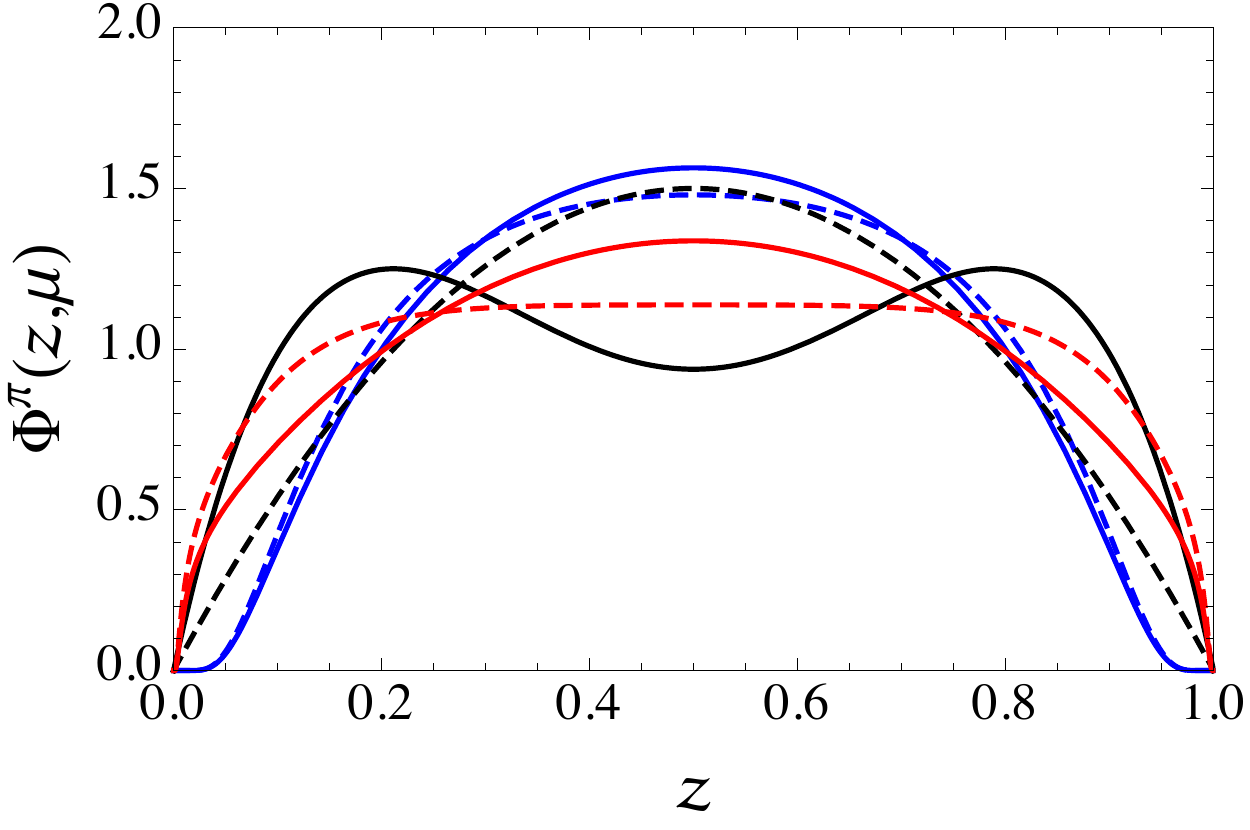}}\quad
\subfigure[]{\includegraphics[scale=0.65]{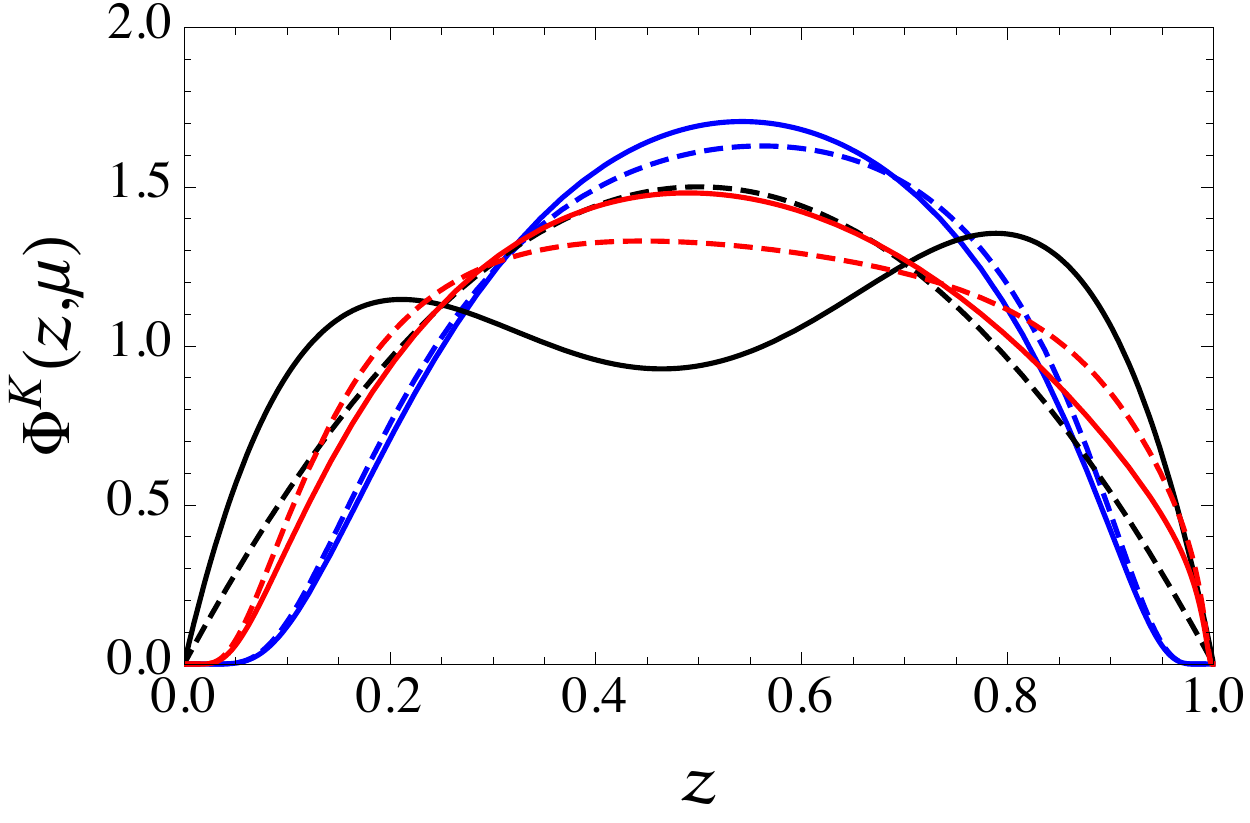}}\\[0.2cm]
\subfigure[]{\includegraphics[scale=0.65]{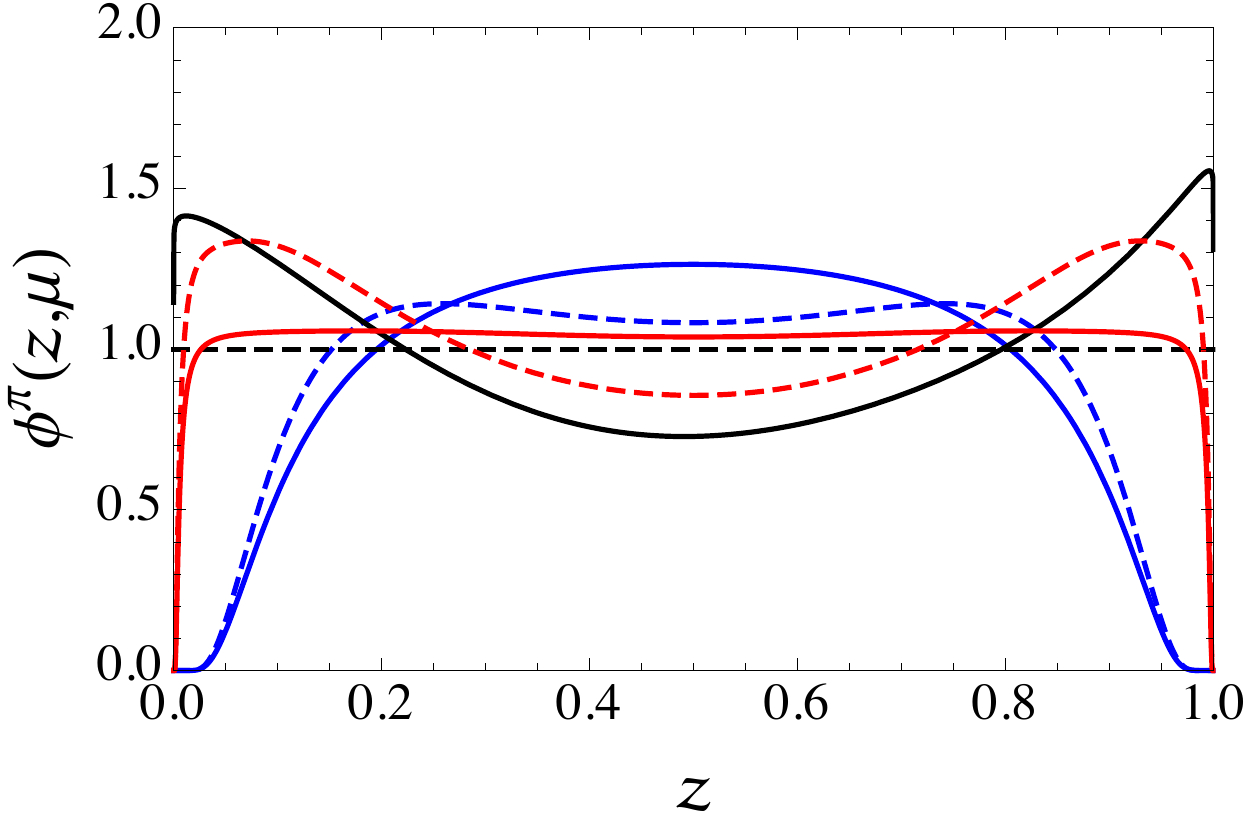}}\quad
\subfigure[]{\includegraphics[scale=0.65]{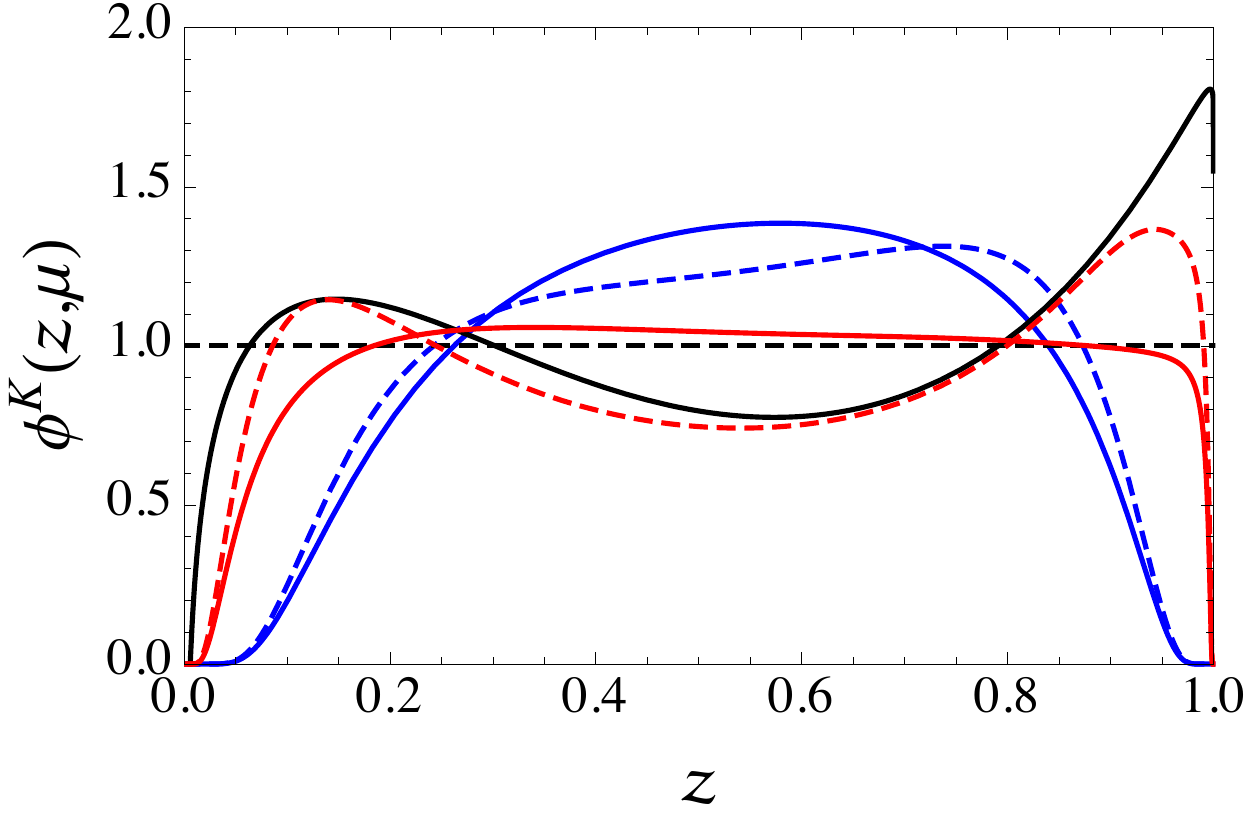}}
\caption{The holographic DAs of $\pi$ and $K$ mesons in S1~(blue) and S2~(red) at $0.5\,{\rm GeV}$~(dashed) and $1\,{\rm GeV}$~(solid), compared with the asymptotic DAs~(black dashed) and the DAs at $1\,{\rm GeV}$ in QCDSR approach~(black solid). \label{fig:DAs}}
\end{center}
\end{figure}

Using the decay constants obtained above and the formulae given in section 2, we now present in Fig.~\ref{fig:DAs} our predictions for the LF holographic DAs of $\pi$ and $K$ mesons at $\mu=1\,{\rm GeV}$ and $0.5\,{\rm GeV}$ in both S1 and S2. For comparison, the asymptotic forms of DAs,
$\Phi(z)=6z\bar{z}$ and $ \phi(z)=1\,$,
and the DAs predicted by QCD sum rule~(QCDSR) approach~\cite{Ball:2006wn} (see Ref.~\cite{Ball:2006wn} for detail) are also plotted in Fig.~\ref{fig:DAs}.

Using the normalization factor $N$ determined by the normalization condition for LFWF, Eq.~(\ref{eq:norc}), and the decay constant given by Eqs.~(\ref{eq:dck1}) and (\ref{eq:dck2}), we find that our extracted holographic DA, $\Phi(z,\mu)$, for both S1 and S2 automatically satisfy the normalization condition $\int_0^1\d z\,\Phi(z,\mu)\,=1$. However, the extracted twist-3 holographic DAs only satisfy the normalization condition   approximately. One of the main reasons is that, in contrast to the case of twist-2 DAs, the normalization of twist-3 holographic DAs is affected by the scale-dependent running masses of light quarks, $\bar{m}_{q,\bar{q}}(\mu)$, appearing in $\mu_P$, which have large uncertainties and are not well determined at low scales. In our evaluation, the values $\bar{m}_s(1{\rm GeV})= 128\,{\rm MeV}$ and $\bar{m}_s/\bar{m}_{u,d}= 24$ are used. It should be noted that, in the evaluation of hadronic matrix elements using the holographic DAs, the effect of $\bar{m}_q(\mu)$ vanishes because the factor $\mu_P$ is cancelled, which can be clearly seen from, for instance, Eqs.~\eqref{eq:twi3a} and \eqref{eq:lfdak3p}. Moreover, we find that the decay constant is also cancelled, as is generally expected since all of the hadronic information is encoded in the LFWFs. All the above findings could also be clearly seen from our following discussions of pure annihilation $\bar{B}_{s}\to \pi^+ \pi^-$ and $\bar{B}_{d}\to K^+K^-$ decays.

From Fig.~\ref{fig:DAs}, comparing the curves of holographic DAs at $\mu=0.5\,{\rm GeV}$ and $1\,{\rm GeV}$ with each other, it can be seen that the effect of evolution is significant at low scale. The evolution at large scale is not obvious, as found also in the previous works~\cite{Ahmady:2012dy,Ahmady:2013cva}, and the perturbative evolution could be in principle recovered through the Efremov-Radyushkin-Brodsky-Lepage (ERBL) equation~\cite{Lepage:1979zb,Efremov:1978rn,Efremov:1979qk}  as has been done in Ref.~\cite{Brodsky:2011yv}.

Comparing the DAs, we can see from Fig.~\ref{fig:DAs} that the twist-2 holographic DA in S2 is considerably broader than the asymptotic form as expected in the other theories such as QCDSR, while the one in S1 is much narrower than in S2. For the twist-3 holographic DA, the behavior in S2 at low scale is similar to the QCDSR result; at large scale it is similar to the asymptotic form except at the regions near end-point; while the twist-3 DA in S1 is suppressed for $z,\bar{z}\lesssim0.2$.

In contrast to  the DAs of asymptotic form and the QCDSR results, the essential feature of LF holographic DAs is that all of the DAs fall rapidly to zero when $z\to 0$ and $1$, which is due to the exponential term, $e^{-\frac{1}{2\lambda}(\frac{m_q^2}{x}+\frac{m_{\bar{q}}^2}{1-x} )}$, in the LFWF given by Eq.~(\ref{eq:LFWFkT}). This implies that the light-front holographic DAs could provide a way to regulate the end-point divergence in the annihilation amplitudes of heavy hadron weak decays.

\subsection{The moments and inverse moment}

\begin{table}[t]
\caption{\label{tab:mompi} The (inverse) moments of $\pi^-$ meson at $1\,{\rm GeV}$ while for Refs.~\cite{Chernyak:1983ej,Braun:2006dg} at $2\,{\rm GeV}$.}
\let\oldarraystretch=\arraystretch
\renewcommand*{\arraystretch}{1.1}
\begin{center}\setlength{\tabcolsep}{5.8pt}
\begin{tabular}{lcccccccccccc}
\hline\hline
&S1        &S2       &Asym.   & LFQM  &QCDSR &QCDSR &LQCD & NLCQM &DES & RM\\
& &  &  &  \cite{Choi:2007yu}  & \cite{Ball:2006wn}& \cite{Chernyak:1983ej} &\cite{Braun:2006dg}&  \cite{Nam:2006au}& \cite{Chang:2013pq} & \cite{Agaev:2005rc} \\ \hline
$\la\xi_2\ra$& $0.172$&$0.238$&$0.2$   &$0.24$ &$0.286$  &$0.343$ &$0.269$&$0.21$  &$0.28$&  $0.28$\\
$\la\xi_4\ra$& $0.062$&$0.116$&$0.086$&$0.11$&$0.143$ &$0.181$ &--- &$0.09$  &$0.15$ & $0.13$ \\
$\la z^{-1}\ra$&$2.61$&$3.50$ &$3$       &---       &$3.75$   &$4.25$ &--- &---       &$5.5$     &---\\
\hline\hline
\end{tabular}
\end{center}
\end{table}

\begin{table}[t]
\caption{ The (inverse) moments of $K^-$ meson at $1\,{\rm GeV}$ while for Ref.~\cite{Braun:2006dg} at $2\,{\rm GeV}$. }
\let\oldarraystretch=\arraystretch
\renewcommand*{\arraystretch}{1.1}
\begin{center}\setlength{\tabcolsep}{5.8pt}
\begin{tabular}{lccccccccc}
\hline\hline
&S1        &S2       &Asym.   & LFQM\cite{Choi:2007yu}  & QCDSR \cite{Ball:2006wn} &LQCD~\cite{Braun:2006dg}&NLCQM \cite{Nam:2006au}\\ \hline
$\la\xi_1\ra$& $0.060$&$0.010$&$0$     &$0.06$&$0.036$&--- &$0.057$\\
$\la\xi_2\ra$& $0.155$&$0.212$&$0.2$   &$0.21$&$0.286$&$0.260$ &$0.182$\\
$\la\xi_3\ra$& $0.025$&$0.014$&$0$     &$0.03$&$0.015$&--- &$0.023$\\
$\la\xi_4\ra$& $0.052$&$0.093$&$0.086$&$0.09$&$0.143$&--- &$0.070$ \\
$\la z^{-1}\ra$&$2.28$&$2.79$ &$3$       &---      &$3.57$  &--- &--- \\
\hline\hline
\end{tabular}
\end{center}
\label{tab:momk}
\end{table}

In order to further compare the predictions based on the holographic DAs with the ones from the other non-perturbative methods, we compute the expectation values of the longitudinal momentum, the $\xi$-moments and the inverse moment, which are defined, respectively, by
\begin{eqnarray}
\la\xi_n\ra=\int_0^1\d z \,(2z-1)^n \Phi(z,\mu)\,,\quad \la z^{-1}\ra=\int_0^1\d z\, z^{-1}\Phi(z,\mu)\,.
\end{eqnarray}
Using the central values of input parameters, our numerical results are listed in Tables~\ref{tab:mompi}~(for $\pi^-$) and \ref{tab:momk}~(for $K^-$). The theoretical predictions based on the LF quark model~(LFQM)~\cite{Choi:2007yu}, the QCDSR~\cite{Ball:2006wn,Chernyak:1983ej}, the LQCD~\cite{Braun:2006dg}, the nonlocal chiral quark model~(NLCQM)~\cite{Nam:2006au}, the Dyson-Schwinger equations~(DSE)~\cite{Chang:2013pq} and the renormalon method~(RM)~\cite{Agaev:2005rc} are also summarized in Tables~\ref{tab:mompi} and \ref{tab:momk} for comparison.

Comparing with the predictions for moments in the other theoretical models,  although the results based on the holographic DA in S1 result in a better agreement than the ones without helicity-improvement as found in Ref.~\cite{Ahmady:2016ufq}, they are still very small~(even smaller than the results using the asymptotic DA) which can be seen from  Tables~\ref{tab:mompi} and \ref{tab:momk}. As argued in Ref.~\cite{Ahmady:2016ufq}, such discrepancies might be attributed to the fact that the dynamical spin effects are not fully captured by S1. Fortunately, as exhibited in Tables~\ref{tab:mompi} and \ref{tab:momk}, we find that such discrepancies are eliminated by adopting the holographic DA in S2.

\subsection{Pion-to-photon transition form factor}

\begin{figure}[t]
\begin{center}
\includegraphics[scale=0.55]{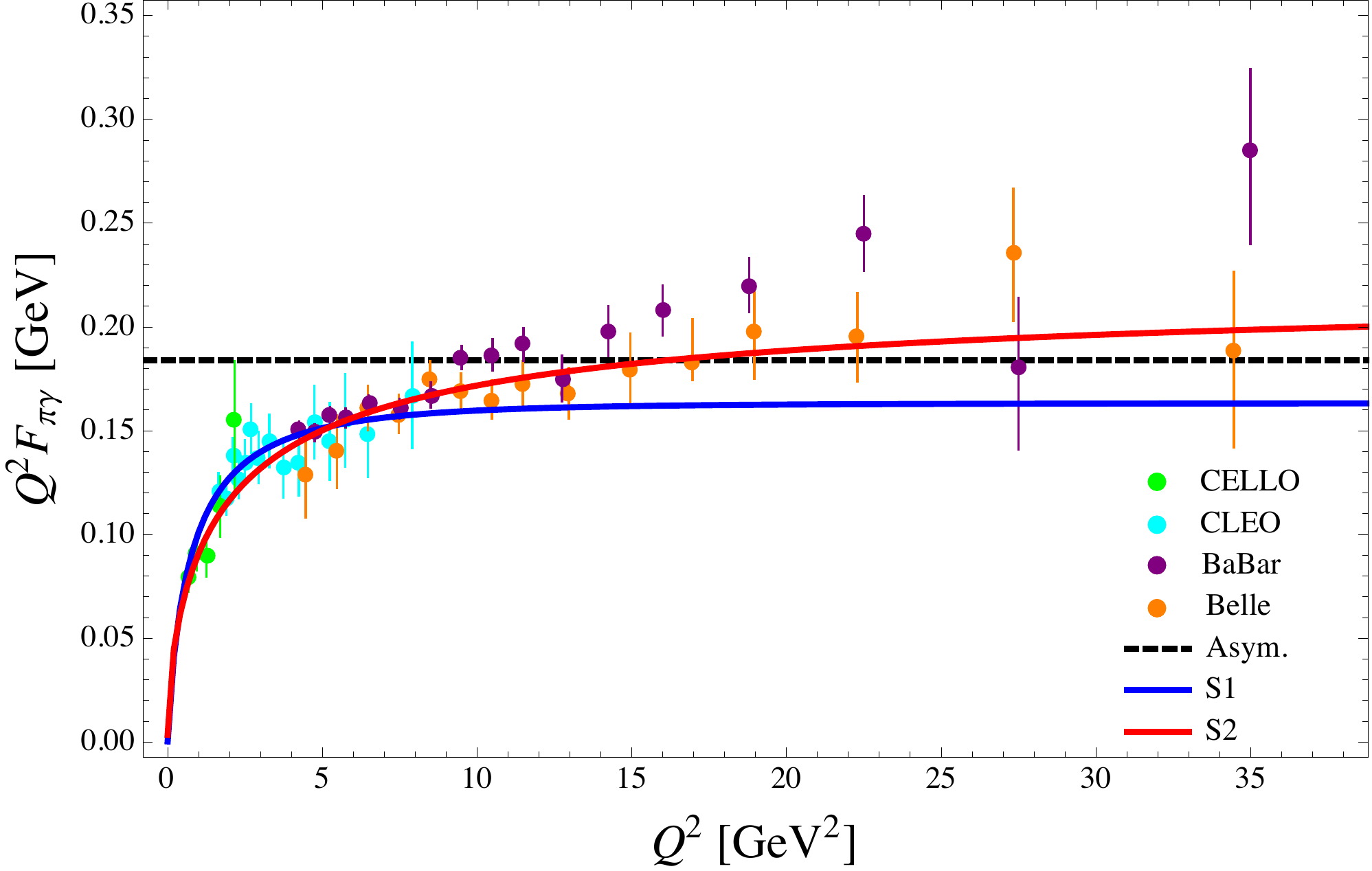}
\caption{\label{pirFF} Theoretical predictions for $Q^2\,F_{\pi\r}(Q^2)$ with asymptotic DA~(black dashed), holographic DAs of S1~(blue) and S2~(red), together with the comparison to the experimental data from CELLO~(green)~\cite{Behrend:1990sr}, CLEO~(cyan)~\cite{Gronberg:1997fj}, BaBar~(Purple)~\cite{Aubert:2009mc} and Belle~(orange)~\cite{Uehara:2012ag}.}
\end{center}
\end{figure}

The pion-to-photon transition form factor can be extracted from the process $\r^*(q_1) \r^*(q_2) \to \pi$. In the case of only one photon being off-shell the transition form factor is denoted as $F_{\pi\r}(Q^2)$ and, to the leading order in $\alpha_s$, is given as~\cite{Lepage:1980fj,Brodsky:2011yv}
\begin{eqnarray}
Q^2\,F_{\pi\r}(Q^2)=\frac{\sqrt{2}}{3}f_{\pi}\int_0^1\d z\,\frac{\Phi^{\pi}(z,zQ)}{z}\,.
\end{eqnarray}

Using the asymptotic DA and the holographic DAs of S1 and S2, the dependence of the rescaled form factor, $Q^2\,F_{\pi\r}(Q^2)$, on the photon virtuality, $Q^2$, are plotted in Fig.~\ref{pirFF}, in which the data from CELLO~\cite{Behrend:1990sr}, CLEO~\cite{Gronberg:1997fj}, BaBar~\cite{Aubert:2009mc} and Belle~\cite{Uehara:2012ag} Collaborations are also shown for comparison. Even though the  holographic DA of S1 does a better job than the traditional one~\cite{Ahmady:2016ufq}, its prediction for $Q^2\,F_{\pi\r}(Q^2)$ is always smaller than the one obtained with asymptotic DA, which is disfavored by the BaBar~\cite{Aubert:2009mc} and Belle~\cite{Uehara:2012ag} data at large $Q^2$ domain. Such an inconsistency could be significantly improved by the holographic DA of S2.  As shown clearly in Fig.~\ref{pirFF}, it can be also found that the holographic DA of S2 does a good job  explaining the current data in the whole $Q^2$ domain, except for the result of  BaBar Collaboration~\footnote{It should be noted that the BaBar and Belle measurements for $Q^2\,F_{\pi\r}(Q^2)$  at large $Q^2$ domain are not consistent with each other.  }.

\section{Pure annihilation $\bar{B}_{s}\to \pi^+ \pi^-$ and $\bar{B}_{d}\to K^+K^-$ decays}
The two-body pure annihilation $B$-meson decays have attracted much theoretical attention during the past years, for instance, in Refs.~\cite{Arnesen:2006dc,Xiao:2011tx,Ali:2007ff,Cheng:2009mu,Chang:2014yma,
Chang:2014rla,Chang:2012xv,Bobeth:2014rra,Li:2015xna,Zhu:2011mm,Wang:2013fya,
Li:2004ep,Yang:2005nra}. The experimental evidence for pure annihilation $\bar{B}_{s}\to \pi^+ \pi^-$ and $\bar{B}_{d}\to K^+K^-$ decays was reported first by the CDF Collaboration~\cite{CDFanni}, and was soon confirmed and updated by both Belle~\cite{Duh:2012ie} and LHCb Collaborations~\cite{LHCbanni,LHCbanni16}.  The Heavy Flavor Averaging Group (HFAG) presents the following averaged results for the branching ratios~\cite{HFAG}:
  \begin{eqnarray}
 {\cal B}(\bar{B}_{s}{\to}{\pi}^{+}{\pi}^{-})
  &=&(6.71{\pm}0.83){\times}10^{-7}\,,
  \label{HFAGpipi}\\[0.2cm]
 {\cal B}(\bar{B}_{d}{\to}K^{+}K^{-})
  &=&(0.84\pm 0.24){\times}10^{-7}
  \label{HFAGKK},
  \end{eqnarray}
with the corresponding significances at the levels of about $5\sigma$ and $3\sigma$, respectively. These measurements require accurate theoretical evaluations. However, due to the appearance of end-point singularities, the annihilation amplitudes cannot be reliably calculated. Motivated by the end-point behavior of the LF holographic DAs, we now try to evaluate the annihilation contributions with  LF holographic DAs and check if the end-point divergence can be properly regulated.

\begin{figure}[t]
\begin{center}
\subfigure[]{\includegraphics[scale=0.23]{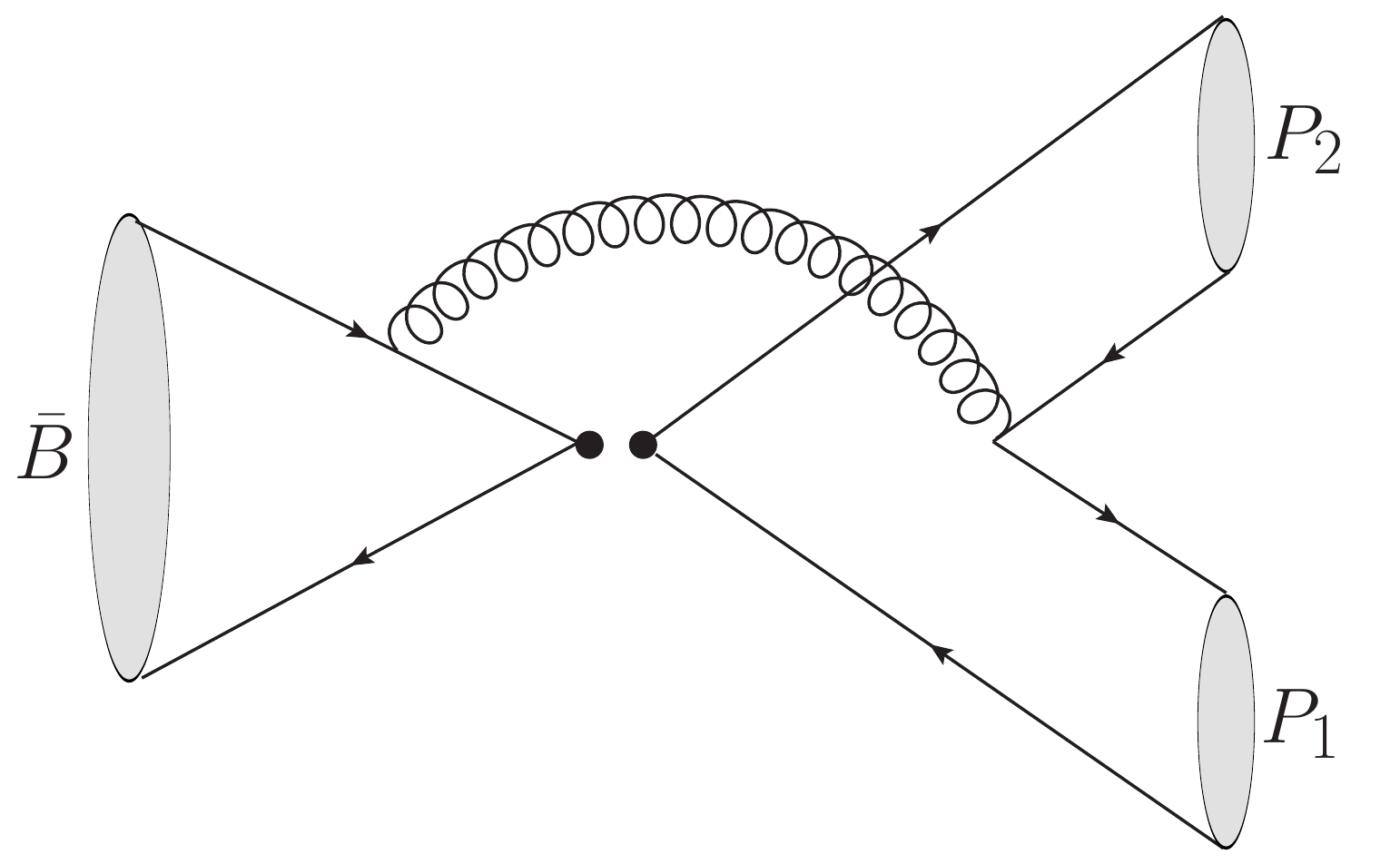}}\quad
\subfigure[]{\includegraphics[scale=0.23]{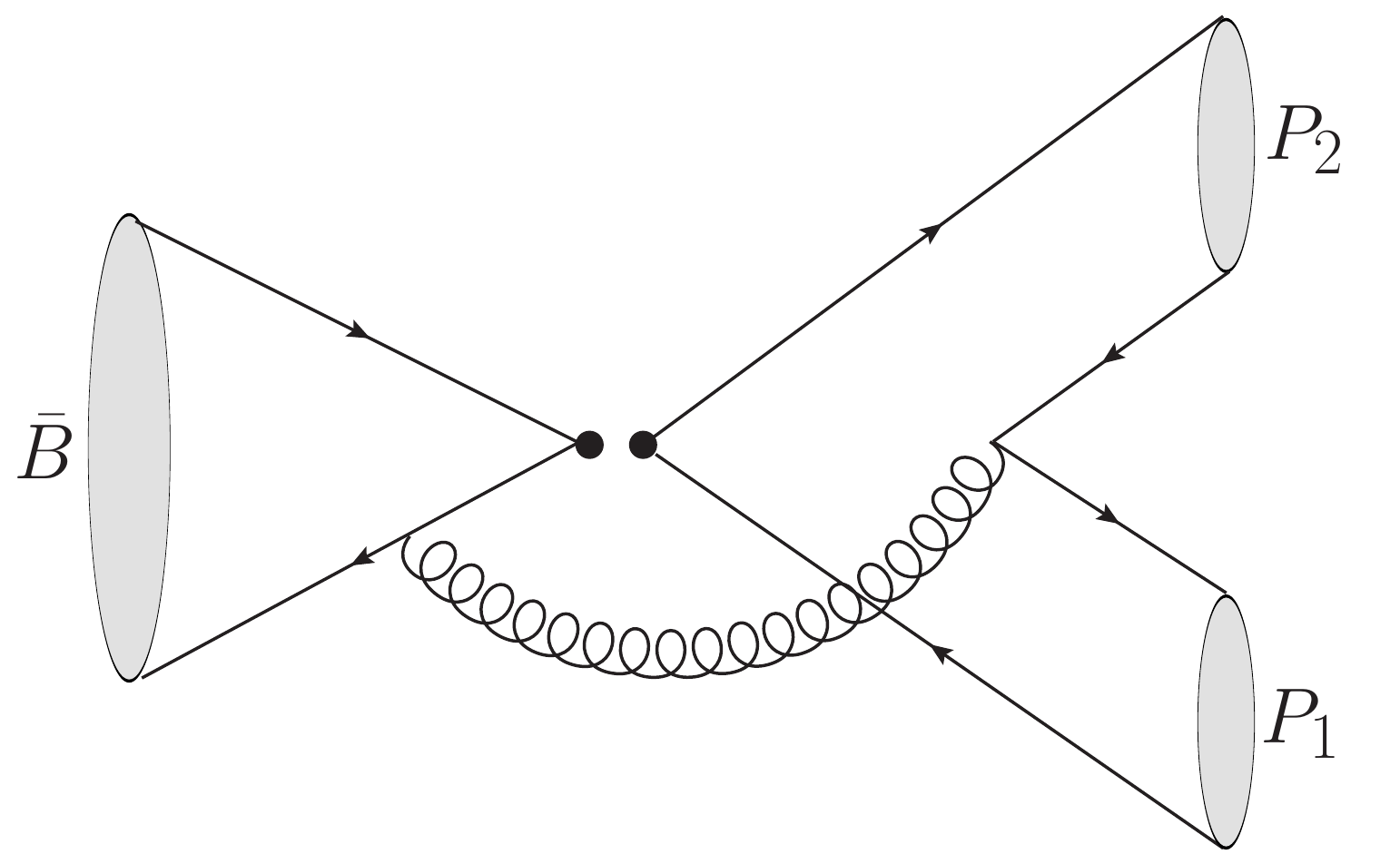}}\quad
\subfigure[]{\includegraphics[scale=0.23]{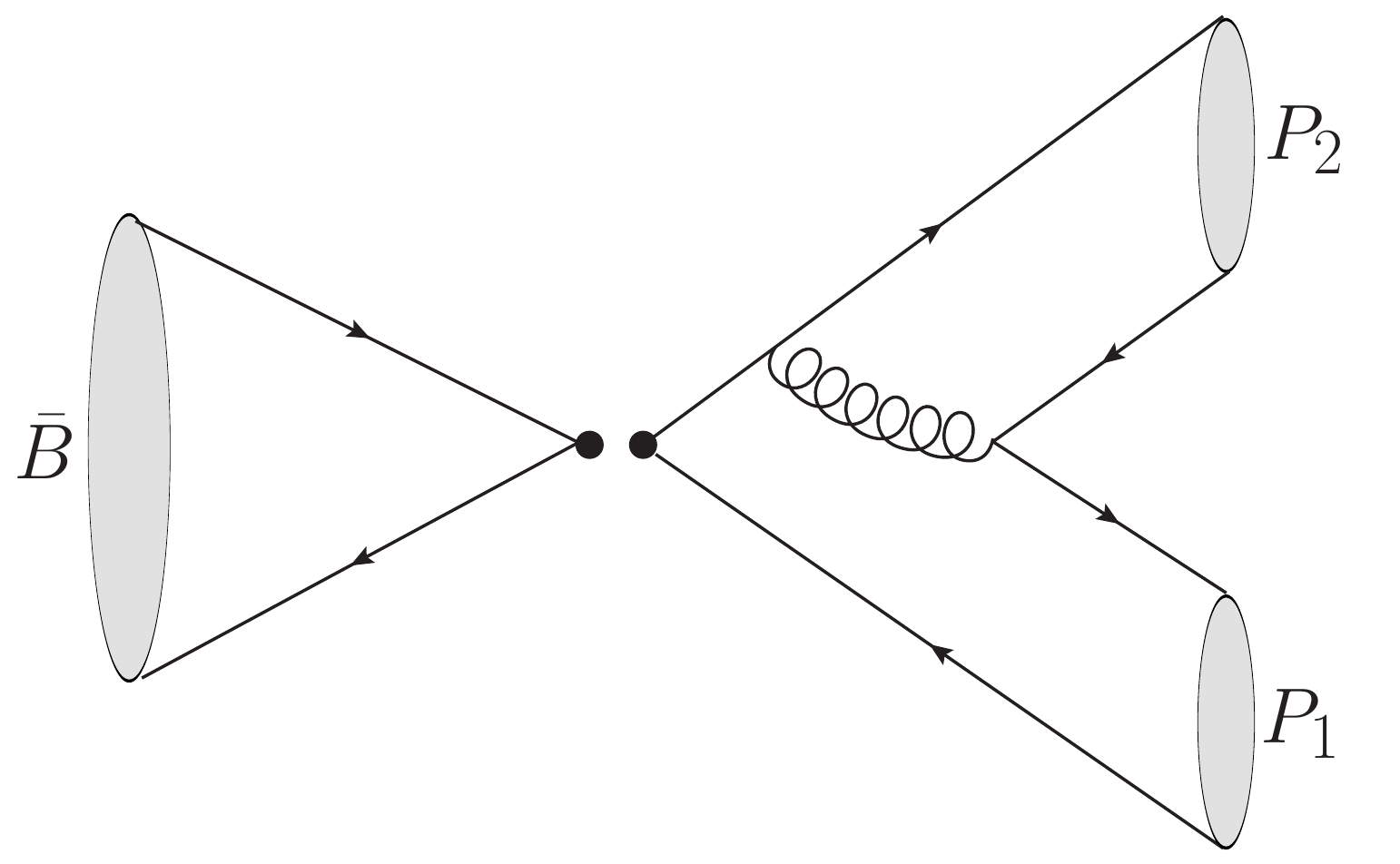}}\quad
\subfigure[]{\includegraphics[scale=0.23]{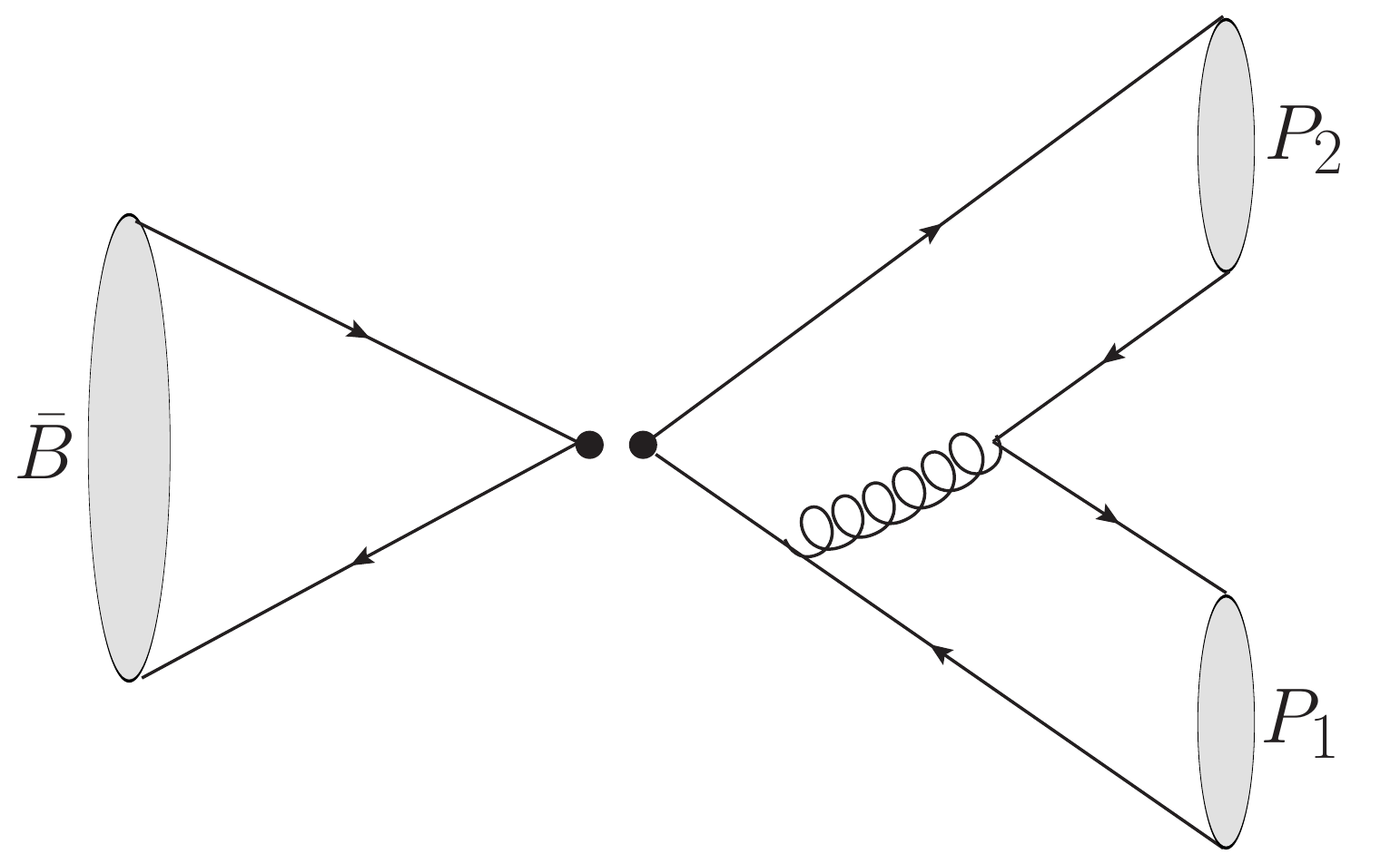}}
\caption{\label{fig:anni} The leading-order Feynman diagrams for pure annihilation $B$-meson decays.}
\end{center}
\end{figure}

Following the prescription proposed in Ref.~\cite{Lepage:1980fj}, the hadronic matrix elements of annihilation topologies can be written as the convolution integrals of the scattering kernel with the DAs of the participating mesons~\cite{Beneke1},
  \begin{eqnarray}
  \langle P_1 P_2|O_i|\bar{B}\rangle =f_B f_{P_1}f_{P_2}\int \d x \d y \d \xi\,{\cal T}_{i}(x,y,\xi)\,\varphi_{P_1}(x)\,\varphi_{P_2}(y)\,\varphi_{B}(\xi)\,,
\label{element}
\end{eqnarray}
where $O_i$ is the local four-quark operator, $x\,,y$ and $\xi $ are (anti-)quark momentum fractions, and the kernel ${\cal T}_{i}(x,y,\xi)$ is obtained by calculating the leading-order Feynman diagrams shown in Fig.~\ref{fig:anni}. In the heavy quark limit and the collinear factorization scheme, the non-zero basic building blocks relevant to $\bar{B}_s\to \pi^+\pi^-$ and $\bar{B}_d\to K^+K^-$ decays can be written as~\cite{Beneke5}
\begin{eqnarray}
 && A_1=
\pi\alpha_s \int_0^1 dx dy
 \bigg\{ \widetilde{\Phi}_{P_2}(x) \widetilde{\Phi}_{P_1}(y)
 \Big[ \frac{1}{y(1-x \bar{y})}
     + \frac{1}{\bar{x}^2 y} \Big]
     +\frac{4}{\bar{m}_b^2(\mu)}
     \frac{2 \widetilde{\phi}_{P_2}(x) \widetilde{\phi}_{P_1}(y)}
          {\bar{x}y} \bigg\}
 \label{ai1}, \\[0.2cm]
&& A_2=
\pi\alpha_s  \int_0^1  dx dy
 \bigg\{ \widetilde{\Phi}_{P_2}(x) \widetilde{\Phi}_{P_1}(y)
 \Big[ \frac{1}{\bar x(1-x \bar{y})}
     + \frac{1}{\bar{x} y^2} \Big]
     + \frac{4}{\bar{m}_b^2(\mu)}
     \frac{2 \widetilde{\phi}_{P_2}(x) \widetilde{\phi}_{P_1}(y)}
     {\bar{x} y} \bigg\}
 \label{ai2},
  \end{eqnarray}
in which,
\begin{equation} \label{eq:rescDAs}
\widetilde{\Phi}_P(z)\equiv f_P{\Phi}_P(z)\,,\qquad\widetilde{\phi}_P(z)\equiv f_P\mu_P{\phi}_P(z)\,,
\end{equation}
and the subscripts $1$ and $2$ correspond to the Dirac current structures of $O_i$, $(V-A)\otimes(V-A)$ and $(V-A)\otimes(V+A)$, respectively. The full amplitudes of $\bar{B}_s\to \pi^+\pi^-$ and $\bar{B}_d\to K^+K^-$ decays are given as
\begin{equation}\label{eq:amp}
 {\cal A}(\bar{B}\to PP)=\sum_{p=u,c}B_{PP}^p\Bigg[\left(\delta_{pu}b_1^p+b_4^p+b_{4,{\rm EW}}^p\right)_{P^-P^+}+\left(b_4^p-\frac{1}{2}b_{4,{\rm EW}}^p\right)_{P^+P^-}\Bigg]\,,
  \end{equation}
with $P^{\pm}=\pi^{\pm}\,,K^{\pm}$ and
\begin{eqnarray}
&& \hspace{2.0cm} B_{\pi\pi}^p=i\frac{G_F}{\sqrt{2}}V_{pb}V_{ps}^*f_{B_s}\,,\qquad B_{KK}^p=i\frac{G_F}{\sqrt{2}}V_{pb}V_{pd}^*f_{B_d}\,,\\[0.2cm]
&& b_{1}^p =
   \frac{C_{F}}{N_{c}^{2}}\, C_{1} A_{1}\,,
   \quad
   b_{4}^{p} =
   \frac{C_{F}}{N_{c}^{2}}\,
   \Big[ C_{4} A_{1} + C_6 A_2\Big]\,,
      \quad
    b_{4,\rm EW}^p =
   \frac{C_{F}}{N_{c}^{2}}\,
   \Big[ C_{10} A_1 + C_8 A_2 \Big]\,,
   \end{eqnarray}
in which, $V_{pb}V_{pd}^*$~($p=u,c$) is the product of the Cabibbo-Kobayashi-Maskawa (CKM) matrix elements~\cite{Cabibbo:1963yz,Kobayashi:1973fv}, and $C_{i}$ the scale-dependent Wilson coefficients. We use the subscripts $P^-P^+$ and $P^+P^-$ in Eq.~\eqref{eq:amp} to indicate that the first meson contains the antiquark emitted from the weak vertex having momentum fraction $\bar{y}$, while another  quark emitted from the weak vertex has momentum fraction $x$.

As mentioned already, using the LF holographic DAs, Eqs.~\eqref{eq:lfdak2} and \eqref{eq:lfdak3p} in S1, and \eqref{eq:lfdak22} and \eqref{eq:lfdak3p2} in S2, one can see that both the decay constants of light mesons and the chiral factor $\mu_P$  in $A_1$ and $A_2$ cancel out. Thus, the hadronic matrix elements do not depend on the decay constants of light mesons and the light-quark running masses when one uses the extracted LF holographic DAs; The hadronic information of light mesons are encoded in the LFWFs.

From Eqs.~\eqref{ai1} and \eqref{ai2}, taking the twist-3 part as an example, one  finds that the end-point divergence  appears when the asymptotic DA, $\phi(z)=1$, or any other forms of DA having non-vanishing end-point behavior are adopted, {\it i.e.},
\begin{eqnarray}
\lim_{\bar{x}\,{\rm or}\,y\to 0} \frac{\phi_{P_2}(x) \phi_{P_1}(y)}{\bar{x} y} = \lim_{\bar{x}\,{\rm or}\,y\to 0} \frac{ 1}{\bar{x}y} =\, \infty\,.
\end{eqnarray}
Traditionally, the corresponding integrals are usually parameterized by a complex parameter $X_A$, according to $\int_0^1dx/x \to X_A = (1+ \rho_A e^{i\phi_A})\ln (m_B/\Lambda_h)$~\cite{Beneke5}. As we have emphasized, in the framework of LF holographic QCD, such an end-point divergence does not exist because it is regulated naturally by the exponential factor in LFWF;
the contributions near the end-point, such as $z\lesssim m_q^2/(2\lambda)$ and $\bar{z}\lesssim m_{\bar{q}}^2/(2\lambda)$ for S2, are suppressed (see Fig. \ref{fig:DAs}).

\begin{table}[t]
\caption{\label{pikbr} The CP-averaged branching ratios of $\bar{B}_s\to \pi^+\pi^-$ and $\bar{B}_d\to K^+K^-$ decays in unit of $10^{-7}$. For the results of S2, the first, second and third theoretical errors are caused by uncertainties of the CKM parameters and $B$-meson decay constants, the holographic parameters in Eq.~\eqref{eq:input2}, and the renormalization scale $\mu$, respectively.}
\let\oldarraystretch=\arraystretch
\renewcommand*{\arraystretch}{1.1}
\begin{center}\setlength{\tabcolsep}{3.5pt}
\begin{tabular}{lccccccc}
  \hline\hline
  Decay Mode & Exp.~\cite{HFAG} & S1&S2 & QCDF~\cite{Beneke5} & pQCD~\cite{Xiao:2011tx}\\ \hline
  $\bar{B}_s \to \pi^+ \pi^-$& $6.71{\pm}0.83$ & $0.220$ & $6.81^{+0.54+1.33+18.41}_{-0.46-1.29-\phantom{0}3.44}$
  & $0.24^{+0.03+0.25+1.63}_{-0.03-0.12-0.21}$& $5.10^{+1.96+0.25+1.05+0.29}_{-1.68-0.19-0.83-0.20}$\\
  $\bar{B}_d \to K^+ K^-$& $0.84\pm 0.24$ & $0.023$ &$0.23^{+0.03+0.06+0.42}_{-0.02-0.06-0.09}$
    & $0.13^{+0.05+0.08+0.87}_{-0.05-0.05-0.11}$&  $1.56^{+0.44+0.23+0.22+0.13}_{-0.42-0.22-0.19-0.09}$\\
  \hline\hline
\end{tabular}
\end{center}
\end{table}

In the numerical evaluations, we will use the values of CKM parameters fitted by the CKMfitter group~\cite{CKMfitter},
\begin{eqnarray}
 A = 0.8227^{+0.0066}_{-0.0136}, \quad
 \lambda = 0.22543^{+0.00042}_{-0.00031},\quad
 \bar{\rho} = 0.1504^{+0.0121}_{-0.0062}, \quad
 \bar{\eta} = 0.3540^{+0.0069}_{-0.0076},
\end{eqnarray}
the averaged values of the $B$-meson decay constants~\cite{PDG},
\begin{eqnarray}
  f_{B_s} = 227.2\pm3.4\,{\rm MeV }\,,\quad f_{B_d} = 190.9\pm 4.1 {\rm MeV }\,,
\end{eqnarray}
and the central values of the other inputs, such as the well-determined masses and lifetimes of $B$ mesons, and the Fermi constant {\it etc.,} given by PDG~\cite{PDG}. Using these inputs, our numerical results for the CP-averaged branching ratios of $\bar{B}_s\to \pi^+\pi^-$ and $\bar{B}_d\to K^+K^-$ decays are listed in Table~\ref{pikbr}, in which the experimental data and the previous theoretical results based on the QCDF with parameterization scheme~\cite{Beneke5} and the pQCD approach~\cite{Xiao:2011tx} are also given for comparison. Our results are evaluated at the renormalization scale $\mu\sim\bar{m}_b/2=2.09\,{\rm GeV }$ with an assigned uncertainty $\pm1\,{\rm GeV }$. For the case of S2, the theoretical errors caused by the CKM parameters and $B$-meson decay constants, the holographic inputs given by Eq.~\eqref{eq:input2}, and the renormalization scale $\mu$ are obtained by evaluating separately the uncertainties induced by each input parameter and then adding them in quadrature.

From Table~\ref{pikbr}, we find that the results in S1 are similar to the central values obtained in QCDF~\cite{Beneke5} with traditional parameterization scheme, but are about one order of magnitude smaller than the data, which is mainly due to the fact that the holographic DAs in S1 are relatively narrow as shown in Fig.~\ref{fig:DAs}, and the contributions with $z$ and $\bar{z}\lesssim 0.2$ are strongly suppressed. In contrast, our prediction for ${\cal B}(\bar{B}_{s}\to\pi^{+}\pi^{-})$ in S2 is in good agreement with the data; within the experimental and theoretical uncertainties our prediction for ${\cal B}(\bar{B}_{d}\to K^{+}K^{-})$ in S2 also agrees with the data. This implies that, compared to S1, S2 is much more favored by the data of ${\cal B}(\bar{B}_{s}\to\pi^{+}\pi^{-})$ and ${\cal B}(\bar{B}_{d}\to K^{+}K^{-})$. In the following discussions, we will focus only on the results using S2.

Comparing with the previous evaluations in QCDF with parameterization scheme for the end-point divergence, we find that the theoretical predictions are remarkably improved by using the holographic DAs. Moreover, the predicting power is retained. Comparing our predictions with the ones in pQCD, we find good agreement for ${\cal B}(\bar{B}_{s}\to\pi^{+}\pi^{-})$; however, our result for ${\cal B}(\bar{B}_{d}\to K^{+}K^{-})$ is smaller than that obtained in pQCD. The significant difference between ${\cal B}(\bar{B}_{s}\to\pi^{+}\pi^{-})$ and ${\cal B}(\bar{B}_{d}\to K^{+}K^{-})$ in our evaluation is easily understandable due to the following  facts:
\begin{enumerate}
\item[(i)] For the $\bar{B}_s\to \pi^+\pi^-$ decay, because
  $|V_{ub}V_{us}^*|\sim |A\lambda^4(\rho-i\eta)| \ll |V_{cb}V_{cs}^*|\sim A\lambda^2$, its decay amplitude, Eq.~\eqref{eq:amp}, can be simplified as
  \begin{eqnarray}\label{eq:simamppipi}
  {\cal A}(\bar{B}_s\to \pi^+\pi^-)\sim B_{\pi\pi}^c 2\,(b_4^c)_{\pi^-\pi^+} \,,
  \end{eqnarray}
  in which $(b_4^c)_{\pi^-\pi^+}=(b_4^c)_{\pi^+\pi^-}$ because the $u$- and $d$-quark difference is not distinguished in this paper. For the $\bar{B}_d\to K^+K^-$ decay, on the other hand, its amplitude can be simplified as
  \begin{eqnarray}\label{eq:simampkk}
  {\cal A}(\bar{B}_d\to K^+K^-)\sim B_{KK}^u\left(b_1^u\right)_{K^-K^+}+B_{KK}^c\Big[\left(b_4^c\right)_{K^-K^+}+ (b_4^c)_{K^+K^-}\Big]\,.
  \end{eqnarray}
  Comparing with Eq.~\eqref{eq:simamppipi}, one can easily find that the first and second terms  in Eq.~\eqref{eq:simampkk} are relatively suppressed by additional Cabibbo factors $\lambda\sim 0.2$ and $\lambda^2\sim 0.048$,  respectively.  Thus, a large ratio $R_{\pi/K}={\cal B}(\bar{B}_{s}\to\pi^{+}\pi^{-})/{\cal B}(\bar{B}_{d}\to K^{+}K^{-})$ is generally expected.
  
\item[(ii)] Moreover,  for the $K^{-(+)}$ meson, as shown by Fig.~\ref{fig:DAs},  the holographic DAs  near the end-point where the (anti-)strange quark carries small momentum fraction is suppressed due to $m_s>m_{u,d}$. As a result, both twist-2 and twist-3 contributions are relatively suppressed for the $\bar{B}_d\to K^+K^-$ decay compared to the $\bar{B}_s\to \pi^+\pi^-$ decay.
In addition, since $f_{B_s}>f_{B_d}$ and the phase space of $\bar{B}_{s}\to \pi^{+}\pi^{-}$ decay is larger than that of $\bar{B}_{d}\to K^{+}K^{-}$ decay, the ratio $R_{\pi/K}$ is further enhanced.
\end{enumerate}
 It should be noted that our evaluations are performed at  leading order and the theoretical uncertainties, especially the one induced by the renormalization scale,  are still quite large. Moreover, the refined  measurements, especially for the $\bar{B}_{d}\to K^{+}K^{-}$ decay, are required for a definite conclusion.

From the phenomenological point of view, an annihilation amplitude with a large strong phase is generally welcome in order to fit experimental data and to explain some puzzles observed in $B$-meson decays~\cite{Chang:2014rla,Bobeth:2014rra,Chang:2014yma,Li:2015xna,Zhu:2011mm,Wang:2013fya}. As a result, a complex parameter $X_A$ has been introduced in the traditional parameterization scheme within the framework of QCDF~\cite{Beneke5}. By using the dynamical gluon mass $m_g(q^2)$ in QCDF approach~\cite{Chang:2012xv} or by introducing transverse momentum $k_{T}$ degree in pQCD approach~\cite{KLS1,KLS2,Lu:2000em}, a large imaginary part in the annihilation amplitudes is also obtained because the singularities exist in the integral over momentum fractions. In contrast to the above regulation schemes,  the leading-order annihilation contributions are real by using the holographic DAs. This result is understandable due to the fact that, although the leading-order annihilation corrections are evaluated at the order $\alpha_s$, they are in fact ``tree" contributions and there is no independent internal momentum; while, the strong phases are generally induced by the loop integration, such as in the vertex and penguin diagrams.  In the SCET approach,  real annihilation contributions for the leading terms of order ${\cal O}(\alpha_s(m_b)\Lambda_{\rm QCD}/m_b)$ have also been predicted~\cite{Arnesen:2006vb}. In addition, it should be noted that complex annihilation contributions are of course possible if, for instance,  final-state interactions or  higher-order corrections are taken into account.

\section{Summary}

Motivated by the rapid development of the LF holographic QCD, the LFWFs for light pseudoscalar mesons and their applications are studied in this paper. In order to restore the dynamical spin effects of quarks and to improve the predictability of LFWFs for different pseudoscalar mesons, the traditional LFWFs are modified according to two assumptions for the helicity-dependent  wavefunctions, corresponding to the structures $\bar{u}_{h}(i\r_5)v_{\bar{h}}$~(named as S1) and $\bar{u}_{h}(\frac{\widetilde{m}_P}{2p^+}i\r^+\r_5+i\r_5)v_{\bar{h}}$~(named as S2), respectively. The LF holographic DAs of  pseudoscalar mesons are then extracted using the helicity-improved LFWFs. The decay constants, the  $\xi$-moments, the pion-to-photon transition form factor and the $\bar{B}_{s}\to \pi^+ \pi^-$ and $\bar{B}_{d}\to K^+K^-$ decays are then evaluated and compared  with experiment. Our main findings are summarized as follows:
\begin{itemize}
 \item In contrast to the LFWF for S1, we find that the LFWF for S2 can provide sufficient flavor-asymmetry resources for predicting the decay constants of $\pi$ and $K$ mesons. Moreover, the results based on S2 for all of the observables considered in this paper are in a much better agreement with experiment than the ones based on S1.

 \item Taking the decay constants of $\pi$ and $K$ mesons as constraints, we perform a $\chi^2$-fit for the holographic parameters, $\sqrt{\lambda}$ and effective quark masses $m_{u,d}$ and $m_s$. Interestingly, our fitted results are remarkably consistent with the ones obtained by fitting the Regge trajectory of light-quark pseudoscalar mesons.

 \item A new scheme with LF holographic DAs for regulating the end-point divergence in the annihilation amplitudes of $B\to PP$ decays is presented. In this scheme, the leading-order annihilation contributions are real. Numerically, our predictions for the branching fractions ${\cal B}(\bar{B}_{s}\to\pi^{+}\pi^{-})$ and ${\cal B}(\bar{B}_{d}\to K^{+}K^{-})$ using the LF holographic DAs in S2 agree well with current data and result in a relatively large flavor-symmetry breaking effect. These predictions will be further tested by future refined measurements.
\end{itemize}

\section*{Acknowledgements}

We thank Junfeng Sun at HNNU and Xingbo Yuan at KIAST for helpful discussions. This work is supported by the National Natural Science Foundation of China (Grant Nos. 11475055, 11675061 and 11435003). Q.~Chang is also supported by the Foundation for the Author of National Excellent Doctoral Dissertation of P.~R.~China (Grant No. 201317), the Program for Science and Technology Innovation Talents in Universities of Henan Province (Grant No. 14HASTIT036), the Excellent Youth Foundation of HNNU and the CSC (Grant No. 201508410213).  X.~Li is also supported in part by the SRF for ROCS, SEM, and by the self-determined research funds of CCNU from the colleges' basic research and operation of MOE (CCNU15A02037). S.J.B. is supported by the
Department of Energy, contract  DE-AC02-76SF00515. SLAC-PUB-16849.


\end{document}